\documentclass{SciPost}

\usepackage{tikz}
\usetikzlibrary{arrows,arrows.meta,shapes,patterns,decorations.markings}

\usepackage[normalem]{ulem}

\usepackage{subcaption,caption,cases}

\hypersetup{
    colorlinks,
    linkcolor={red!50!black},
    citecolor={blue!50!black},
    urlcolor={blue!80!black}
}

\usepackage[bitstream-charter]{mathdesign}
\DeclareSymbolFont{usualmathcal}{OMS}{cmsy}{m}{n}
\DeclareSymbolFontAlphabet{\mathcal}{usualmathcal}

\fancypagestyle{SPstyle}{
\fancyhf{}
\lhead{\colorbox{scipostblue}{\bf \color{white} ~SciPost Physics }}
\rhead{{\bf \color{scipostdeepblue} ~Submission }}

\fancyfoot[C]{\textbf{\thepage}}
}

\begin{document}

\pagestyle{SPstyle}

\begin{center}{\Large \textbf{\color{scipostdeepblue}{A bottom-up approach to fluctuating hydrodynamics: Coarse-graining of stochastic lattice gases and the Dean-Kawasaki equation\\
}}}\end{center}

\begin{center}\textbf{
Soumyabrata Saha\textsuperscript{1$\star$},
Sandeep Jangid\textsuperscript{1$\dagger$},
Thibaut Arnoulx de Pirey\textsuperscript{2$\ddagger$},
Juliane U. Klamser\textsuperscript{3$\circ$}, and
Tridib Sadhu\textsuperscript{1$\S$}
}\end{center}

\begin{center}
{\bf 1} Department of Theoretical Physics, Tata Institute of Fundamental Research, Homi Bhabha Road, Mumbai 400005, India
\\
{\bf 2} Institut de Physique Théorique, CEA Saclay and CNRS (UMR 3681), Gif-sur-Yvette, 91190, France
\\
{\bf 3} Laboratoire Charles Coulomb (L2C), Université de Montpellier and CNRS (UMR 5221), 34095 Montpellier, France
\\[\baselineskip]
$\star$ \href{mailto:soumyabrata.saha@tifr.res.in}{\small soumyabrata.saha@tifr.res.in}\,,\quad
$\dagger$ \href{mailto:sandeep.jangid@tifr.res.in}{\small sandeep.jangid@tifr.res.in}\,,\quad
$\ddagger$ \href{mailto:thibaut.arnoulxdepirey@ipht.fr}{\small thibaut.arnoulxdepirey@ipht.fr}\,,\quad
$\circ$ \href{mailto:juliane.klamser@umontpellier.fr}{\small juliane.klamser@umontpellier.fr}\,,\quad
$\S$ \href{mailto:tridib@theory.tifr.res.in}{\small tridib@theory.tifr.res.in}
\end{center}

\section*{\color{scipostdeepblue}{Abstract}}
\textbf{\boldmath{Fluctuating hydrodynamics provides a quantitative, large‑scale description of many‑body systems in terms of smooth variables, with microscopic details entering only through a small set of transport coefficients. Although this framework has been highly successful in characterizing macroscopic fluctuations and correlations, a systematic derivation of fluctuating hydrodynamics from underlying stochastic microscopic dynamics remains obscure for broad classes of interacting systems. For stochastic lattice‑gas models with gradient dynamics and a single conserved density, we develop a path‑integral based coarse‑graining procedure that recovers fluctuating hydrodynamics in a controlled manner. Our analysis highlights the essential role of local‑equilibrium averages, which go beyond na\"ive mean‑field–type gradient expansions. We further extend this approach to interacting Brownian particles by coarse‑graining the Dean–Kawasaki equation, revealing a mobility proportional to the density and a diffusivity determined by the thermodynamic pressure.
}}

\vspace{\baselineskip}

\noindent\textcolor{white!90!black}{%
\fbox{\parbox{0.975\linewidth}{%
\textcolor{white!40!black}{\begin{tabular}{lr}%
  \begin{minipage}{0.6\textwidth}%
    {\small Copyright attribution to authors. \newline
    This work is a submission to SciPost Physics. \newline
    License information to appear upon publication. \newline
    Publication information to appear upon publication.}
  \end{minipage} & \begin{minipage}{0.4\textwidth}
    {\small Received Date \newline Accepted Date \newline Published Date}%
  \end{minipage}
\end{tabular}}
}}
}


\vspace{10pt}
\noindent\rule{\textwidth}{1pt}
\tableofcontents
\noindent\rule{\textwidth}{1pt}
\vspace{10pt}

\section{Introduction} \label{sec:intro}

Hydrodynamics is a coarse-grained description of many-particle systems at a macroscopic scale, formulated in terms of a few slow degrees of freedom \cite{1980_Lifshitz_Statistical,1991_Spohn_Large,1999_Kipnis_Scaling,2006_Zarate_Hydrodynamic,2013_Marchetti_Hydrodynamics,2020_Gromov_Fracton,2024_Wienand_Emergence,2025_Doyon_Generalized}, compared to thermodynamically many microscopic variables. These slow variables relate to local conservation laws, such as those of mass, momentum, and energy, which constrain the microscopic dynamics such that their relaxation occurs over large length and time scales. The hydrodynamic description expresses the evolution of these slow conserved variables through classical partial differential equations, obtained by combining continuity equations with constitutive relations that encode transport properties. This coarse-grained description captures the average dynamics of extended many-particle systems at large length and long times without explicit reference to precise microscopic details.

The deterministic description of classical hydrodynamics, however, does not tell the full story as residual fluctuations persist even in large systems and over long times. Coarse-graining the microscopic dynamics by integrating out the fast degrees of freedom leaves a stochastic imprint on the slow conserved modes. Though typically small compared to the mean, these fluctuations become particularly relevant in non-equilibrium settings \cite{2001_Bertini_Fluctuations,2002_Bertini_Macroscopic,2007_Derrida_Non,2015_Bertini_Macroscopic} and play a crucial role in generating long-range spatio-temporal correlations \cite{1990_Garrido_Long,2008_Bodineau_Long,2009_Bertini_Towards,2015_Krapivsky_Dynamical,2015_Sadhu_Large}, nonlocal response to perturbations \cite{2016_Sadhu_Correlations}, anomalous transport \cite{2014_Krapivsky_Large,2015_Krapivsky_Tagged}, rare-event phenomena \cite{2024_Saha_Large,2025_Sharma_Large,2025_Saha_Large}, and dynamical phase transitions \cite{2005_Bodineau_Distribution,2012_Bunin_Non,2013_Bunin_Cusp,2015_Bertini_Macroscopic}. To account for them, classical hydrodynamics must be extended to include a consistent stochastic description, giving rise to \emph{fluctuating hydrodynamics}.

The conventional approach for classical hydrodynamics \cite{1991_Spohn_Large,1999_Kipnis_Scaling} already illustrates the foundations of fluctuating hydrodynamics. The simplest route \cite{2007_Derrida_Non,2009_Derrida_Current2,2011_Derrida_Microscopic} is to combine local conservation laws with phenomenological constitutive relations, such as Fick’s law for diffusion or Fourier’s law for heat transport. A more systematic derivation comes from kinetic theory, where one starts from the Boltzmann equation \cite{2019_Miron_Derivation,2022_Gaspard_The} and extracts macroscopic transport equations and coefficients through controlled expansions. Projection-operator methods \cite{2001_Zwanzig_Non} provide yet another formalism by formally separating the slow conserved modes from the fast microscopic ones. Most earlier discussions \cite{1990_Garrido_Long,1991_Spohn_Large,2014_Spohn_Nonlinear} along this direction are confined to small typical fluctuations around classical hydrodynamics.

Fluctuating hydrodynamics extends these early descriptions by supplementing deterministic conservation laws with \emph{multiplicative} noise terms that describe even large fluctuations of the slow conserved fields around their mean behavior. The structure of these noise terms is fixed by the underlying symmetries and conservation laws, while their strength is determined by equilibrium response functions. A natural perspective on their origin comes from an intermediate, mesoscopic description: the system is partitioned into cells large enough to contain many degrees of freedom yet small enough to disregard fluctuations. The currents of conserved quantities exchanged between cells then become stochastic variables. This leads to stochastic conservation laws whose deterministic part reduces to classical hydrodynamics, while the noisy component encodes the residual fluctuations that persist at this scale. 

In this way, fluctuating hydrodynamics serves as a bridge between microscopic models of many-particle systems and the emergent behavior observed on macroscopic scales. It not only characterizes the average transport laws of slow conserved modes but also provides a consistent statistical description of their fluctuations, making it a powerful framework for the study of a broad class of systems both in and out of equilibrium. This description constitutes the essence of the Macroscopic Fluctuation Theory (MFT) \cite{2001_Bertini_Fluctuations,2002_Bertini_Macroscopic,2007_Derrida_Non,2011_Derrida_Microscopic,2015_Bertini_Macroscopic} that was formulated for large deviations studies of non-equilibrium dynamics.

Despite immense activities in past decades, a genuinely bottom-up derivation of fluctuating hydrodynamics from the underlying stochastic microscopic dynamics remains largely obscure \cite{2013_Das_Coarse,2017_Duran_General,2007_Lefevre_Dynamics,2008_Tailleur_Mapping}. In this \emph{Article}, we propose a systematic coarse-graining method that reproduces quantitative fluctuating hydrodynamic description \cite{2007_Derrida_Non,2015_Bertini_Macroscopic} for the macroscopic evolution of slow conserved fields in extended stochastic many-particle systems. Our coarse-graining method, relying on Physics toolbox, is conceptually simple, broadly applicable to a wide range of many-particle systems, and, crucially, provides precise quantitative predictions for atypical fluctuations, particularly large deviations.

In this \emph{Article}, we focus on the simplest setting of a class of diffusive systems that conserves a single quantity locally. Let $\rho(x,t)$ denote the coarse-grained density field of the conserved quantity. Conservation requires that $\rho$ obeys a continuity equation, with the associated current consisting of a deterministic diffusive part and a stochastic part. Incorporating both contributions leads to the stochastic diffusion equation describing the evolution of the density field
\begin{equation}\label{eq:fhd_general}
\partial_t\rho=\partial_x\!\left(D(\rho)\partial_x\rho\right)+\frac{1}{\sqrt{\ell}}\partial_x\!\left(\sqrt{\sigma(\rho)}\eta\right) \,,
\end{equation}
where the two transport coefficients, the diffusivity, $D(\rho)$ and the mobility, $\sigma(\rho)$, encode all the relevant microscopic details \cite{2007_Derrida_Non,2011_Derrida_Microscopic,2012_Krapivsky_Fluctuations,2015_Bertini_Macroscopic}. The parameter $\ell$ here denotes the coarse-graining scale and $\eta(x,t)$ is a zero-mean Gaussian white noise, delta-correlated in space and time with unit strength. In Table \ref{tab:trans_params}, we list the transport coefficients of some well-known interacting particle systems which we reproduce using our coarse-graining method. For time-reversible dynamics in the bulk, the transport coefficients are related by fluctuation dissipation relation $2D(\rho)=\sigma(\rho)f''(\rho)$, where $f(\rho)$ is the canonical free energy density \cite{2007_Derrida_Non,2012_Krapivsky_Fluctuations} (inverse temperature $\beta$ is absorbed in the definition). This draws equivalence of \eqref{eq:fhd_general} to the stochastic Model B in the Hohenberg–Halperin classification \cite{1977_Hohenberg_Theory},
\begin{equation}\label{eq:model_b}
\partial_t\rho=\partial_x\!\left(\frac{\sigma(\rho)}{2}\partial_x\frac{\mathrm{d}f(\rho)}{\mathrm{d}\rho}\right)+\frac{1}{\sqrt{\ell}}\partial_x\!\left(\sqrt{\sigma(\rho)}\eta\right).
\end{equation}
Note that $f''(\rho)=\beta\rho^{-1}P'(\rho)$ relates transport coefficients to thermodynamic pressure $P(\rho)$ and equivalently $f''(\rho)=\beta\rho^{-2}\kappa^{-1}$ to the isothermal compressibility, $\kappa$ \cite{1991_Spohn_Large,2014_Krapivsky_Large,2015_Krapivsky_Tagged,2025_Grabsch_Exact}.

In this \emph{Article}, we present a systematic derivation of \eqref{eq:fhd_general} for a class of stochastic lattice gas and overdamped Langevin particles with short-range inter-particle interactions. For the latter, our discussion is about coarse-graining the associated Dean-Kawasaki equation \cite{1994_Kawasaki_Stochastic,1996_Dean_Langevin,2025_Illien_the}. 
Our analysis utilizes a path-integral representation \cite{2022_Pirey_Path} of the stochastic partial differential equation \eqref{eq:fhd_general} using the Martin-Siggia-Rose-Janssen-De Dominicis (MSRJD) formalism \cite{1973_Martin_Statistical,1976_Janssen_On,1976_Dominicis_Techniques,1978_Dominicis_Field}. In this representation, the transition amplitude from an initial density profile $\rho(x,0)$ to a final one $\rho(x,T)$ following \eqref{eq:fhd_general} is
\begin{subequations}\label{eq:trans_prob_general}
\begin{equation}\label{eq:hydro_path_int}
\Pr\left(\rho(x,T)\middle|\rho(x,0)\right)=\int_{\rho(x,0)}^{\rho(x,T)}\left[\mathcal{D}
\widehat{\rho}\right]\left[\mathcal{D}\rho\right]\exp{\left\{-\ell\int_{0}^{T}\mathrm{d}t\left[\int\mathrm{d}x\left(\widehat{\rho}\partial_t\rho\right)-H\left[\rho,\widehat{\rho}\right]\right]\right\}} \,,
\end{equation}
where $\widehat{\rho}$ is the response field and the effective Hamiltonian
\begin{equation}\label{eq:hydro_hamiltonian}
H\left[\rho,\widehat{\rho}\right]=\int\mathrm{d}x\left[\partial_x\widehat{\rho}\left(\frac{\sigma(\rho)}{2}\partial_x\widehat{\rho}-D(\rho)\partial_x\rho\right)\right] \,.
\end{equation}
\end{subequations}
The exponential weight in \eqref{eq:hydro_path_int} defines the `hydrodynamic action'. In writing \eqref{eq:trans_prob_general} we follow the It\^o convention.

Our coarse-graining method derives the path probability \eqref{eq:hydro_path_int} starting from the microscopic dynamics. This requires constructing a similar path integral representation of microscopic variables \cite{2024_Saha_Large,2025_Mukherjee_Hydrodynamics} (see also \cite{2006_Andreanov_Exact,2007_Lefevre_Dynamics,2025_Neville_Hydrodynamic}) and averaging the corresponding path probability over a local equilibrium measure where microscopic variables are locally in equilibrium with respect to a smoothly varying average field $\rho(x,t)$. This local equilibrium averaging of microscopic variables \cite{1991_Spohn_Large} is an essential step that extends the coarse-graining beyond mere gradient-expansions.

We explicitly demonstrate our method for a class of exclusion processes \cite{1968_Macdonald_Kinetics,1969_Macdonald_Concerning,1994_Schutz_Non} in which particles perform symmetric nearest-neighbor hopping subject to extended hard-core or soft repulsive interactions, highlighting the crucial role played by the gradient-structure of the hopping rates \cite{2025_Derrida_Lectures}. We also show that the same coarse-graining method applies to systems of interacting Brownian particles described on microscopic scales by the Dean-Kawasaki equation \cite{1996_Dean_Langevin}. The coarse graining demonstrates how inter-particle interactions in the Dean-Kawasaki equation re-normalizes the diffusivity in the fluctuating hydrodynamics description. This allows us to recover fluctuating hydrodynamics equations \cite{2014_Krapivsky_Large,2015_Krapivsky_Tagged,2025_Illien_the,2025_Grabsch_Exact} that were previously argued using the fluctuation-dissipation relation.

We organize our \emph{Article} in the following order. In Section.~\ref{sec:lattice} we introduce the coarse-graining approach for lattice models and subsequently in Section.~\ref{sec:continuum} we extend the approach for the Dean-Kawasaki equation. In Section.~\ref{sec:numeric_verify} we present numerical confirmation of the fluctuating hydrodynamics for specific model dynamics. The appendix contains a discussion for general dynamics emphasizing the importance of the underlying gradient structure and the appropriate local equilibrium averaging.

\begin{table}[t]
\centering
\begin{tabular}{|c|c|c|}
\hline
The microscopic model & Diffusivity, $D(\rho)$ & Mobility, $\sigma(\rho)$ \\
\hline
Zero Range Process & $g'(\rho)$ & $2g(\rho)$\\[10pt]
Symmetric Simple Exclusion Process & $1$ & $2\rho(1-\rho)$\\[10pt]
Symmetric Simple Multiple Exclusion Process & $\displaystyle{\frac{1}{[1-(M-1)\rho]^2}}$ & $\displaystyle{\frac{2\rho(1-M\rho)}{1-(M-1)\rho}}$\\[10pt]
Symmetric Simple Partial Exclusion Process & $\displaystyle{N}$ & $\displaystyle{2\rho(N-\rho)}$\\[10pt]
Symmetric Simple Inclusion Process & $K$ & $2\rho(K+\rho)$\\[10pt]
Kipnis-Marchioro-Presutti Model & $1$ & $2\rho^2$\\[10pt]
Brownian Hard Rods & $\displaystyle{\frac{1}{(1-a\rho)^2}}$ & $2\rho$\\[10pt]
Short-range interacting Brownian particles & $\displaystyle{\beta\frac{\mathrm{d}P(\rho)}{\mathrm{d}\rho}}$ & $2\rho$\\
\hline
\end{tabular}
\caption{Transport coefficients for different models of diffusive transport recovered using our coarse-graining method. The Symmetric Simple Multiple \cite{1968_Macdonald_Kinetics,2004_Schonherr_Exclusion} and Partial \cite{1994_Schutz_Non,2008_Tailleur_Mapping} Exclusion Processes generalizes the paradigmatic model of the SSEP. Setting $M=1$ and $N=1$ respectively, reduces the models to the SSEP. The Symmetric Simple Inclusion Process \cite{2007_Giardina_Duality}, on the other hand, is an `attractive' counterpart of the SSEP. The Zero Range Process \cite{1970_Spitzer_Interaction} and the Kipnis-Marchioro-Presutti model of heat conduction \cite{1982_Kipnis_Heat} also allow for an explicit derivation of the fluctuating hydrodynamics using the gradient structure of the microscopic current and known local equilibrium measures \cite{2005_Levine_Zero,2005_Bertini_Large}. \label{tab:trans_params}}
\end{table}

\section{Lattice models: Exclusion processes} \label{sec:lattice}

\subsection{Symmetric Simple Double Exclusion Process}\label{sec:ssdep}

\begin{figure}
\centering
\resizebox{\linewidth}{!}
{
\begin{tikzpicture}
\node [circle, draw=black, thick, fill=white, inner sep=0pt, minimum size=10pt] (0) at (0,0) {};
\node [circle, draw=black, thick, fill=white, inner sep=0pt, minimum size=10pt] (15) at (15,0) {};
\draw[thick] (0) -- (15);
\node [circle, draw=black, thick, fill=white, inner sep=0pt, minimum size=10pt] (1) at (1,0) {};
\node [circle, draw=black, thick, fill=blue, inner sep=0pt, minimum size=10pt] (2) at (2,0) {};
\node [circle, draw=black, thick, fill=white, inner sep=0pt, minimum size=10pt] (3) at (3,0) {};
\node [circle, draw=black, thick, fill=blue, inner sep=0pt, minimum size=10pt] (4) at (4,0) {};
\node [circle, draw=black, thick, fill=white, inner sep=0pt, minimum size=10pt] (5) at (5,0) {};
\node [circle, draw=black, thick, fill=blue, inner sep=0pt, minimum size=10pt] (6) at (6,0) {};
\node [circle, draw=black, thick, fill=white, inner sep=0pt, minimum size=10pt] (7) at (7,0) {};
\node [circle, draw=black, thick, fill=white, inner sep=0pt, minimum size=10pt] (8) at (8,0) {};
\node [circle, draw=black, thick, fill=white, inner sep=0pt, minimum size=10pt] (9) at (9,0) {};
\node [circle, draw=black, thick, fill=blue, inner sep=0pt, minimum size=10pt] (10) at (10,0) {};
\node [circle, draw=black, thick, fill=white, inner sep=0pt, minimum size=10pt] (11) at (11,0) {};
\node [circle, draw=black, thick, fill=white, inner sep=0pt, minimum size=10pt] (12) at (12,0) {};
\node [circle, draw=black, thick, fill=blue, inner sep=0pt, minimum size=10pt] (13) at (13,0) {};
\node [circle, draw=black, thick, fill=white, inner sep=0pt, minimum size=10pt] (14) at (14,0) {};
\draw [-{Triangle[width=2.0mm, length=2.0mm]}, very thick] (2.north) to [out=120,in=45] node [midway, above]{\large$1$} (1.north);
\draw [-{Triangle[width=2.0mm, length=2.0mm]}, very thick] (6.north) to [out=60,in=135] node [midway, above]{\large$1$} (7.north);
\draw [-{Triangle[width=2.0mm, length=2.0mm]}, very thick] (10.north) to [out=120,in=45] node [midway, above]{\large$1$} (9.north);
\draw [-{Triangle[width=2.0mm, length=2.0mm]}, very thick] (10.north) to [out=60,in=135] node [midway, above]{\large$1$} (11.north);
\draw [-{Triangle[width=2.0mm, length=2.0mm]}, very thick] (13.north) to [out=120,in=45] node [midway, above]{\large$1$} (12.north);
\draw [-{Triangle[width=2.0mm, length=2.0mm]}, very thick] (13.north) to [out=60,in=135] node [midway, above]{\large$1$} (14.north);
\draw[ultra thick, dashed] (0) -- (-1,0);
\draw[ultra thick, dashed] (15) -- (16,0);
\end{tikzpicture}
}
\caption{\textbf{SSDEP:} A schematic of the dynamics of the symmetric simple \emph{double} exclusion process ($M=2$) on a one-dimensional lattice. Filled circles denote occupied sites, while open circles represent empty sites. Arrows indicate the allowed particle hops, each occurring with rate~$1$.}
\label{fig:ssdep_scheme}
\end{figure}

We consider an extension of the Symmetric Simple Exclusion Process (SSEP) that accounts for excluded-volume effects of extended hard-core particles on a lattice, known as the Symmetric Simple Double Exclusion Process (SSDEP). In the SSDEP, as in the SSEP, particles obey on-site exclusion. In addition, each particle also blocks its immediate neighboring sites from being occupied. The dynamics is a continuous-time Markov process: at rate $1$, a particle hops to its right or left neighbor, provided that both the target site and the nearest neighbor sites are vacant. A schematic of the dynamics is given in Fig. \ref{fig:ssdep_scheme}.

We denote by $n_i(\tau)$ the occupation number of the $i$-th site at time $\tau$, which, in accordance with the on-site exclusion rule, takes the value $1$ if the site is occupied by a particle and $0$ if it is vacant. Furthermore, we use $Y_i(\tau)$ to denote the net hopping across the bond connecting sites $i$ and $(i+1)$ during the infinitesimal time interval from $\tau$ to $\tau+\mathrm{d}\tau$. For the model under consideration, $Y_i(\tau)$ takes the values $\{\pm1,0\}$ with the respective probabilities given as follows
\begin{subnumcases}
{\label{eq:bulk_hops} Y_i(\tau)=}
1 & with prob. $n_i(\tau)\,\big(1-n_{i+1}(\tau)\big)\,\big(1-n_{i+2}(\tau)\big)\,\mathrm{d}\tau$ \,, \\
-1 & with prob. $n_{i+1}(\tau)\,\big(1-n_i(\tau)\big)\,\big(1-n_{i-1}(\tau)\big)\,\mathrm{d}\tau$ \,, \\
0 & with prob. $1-\big[n_i(\tau)\,\big(1-n_{i+1}(\tau)\big)\,\big(1-n_{i+2}(\tau)\big)$\nonumber\\
& \qquad\qquad\qquad\qquad$+n_{i+1}(\tau)\,\big(1-n_i(\tau)\big)\,\big(1-n_{i-1}(\tau)\big)\big]\,\mathrm{d}\tau$ \,.
\end{subnumcases}
Local particle conservation relates the occupation and hopping variables via the discrete continuity equation
\begin{equation}
n_i(\tau+\mathrm{d}\tau)-n_i(\tau)=Y_{i-1}(\tau)-Y_i(\tau) \,,
\end{equation}
valid for all sites $i$ and times $\tau$.

The exact microscopic configuration of the system at time $\tau$ is specified by the set of the occupation variables $\boldsymbol{n}(\tau)\equiv\big\{\dots\,,\,n_{i-1}(\tau)\,,\,n_i(\tau)\,,\,n_{i+1}(\tau)\,,\,\dots\big\}$ while the evolution is governed by the sequence of hopping variables $\big\{\dots\,,\,Y_{i-1}(\tau)\,,\,Y_i(\tau)\,,\,Y_{i+1}(\tau)\,,\,\dots\big\}$. We denote the initial configuration at $\tau=0$ by $\boldsymbol{n}_\text{ini}\equiv\boldsymbol{n}(0)$ and the final configuration at $\tau=\mathcal{T}$ by
$\boldsymbol{n}_\text{fin}\equiv\boldsymbol{n}(\mathcal{T})$. We follow \cite{2006_Andreanov_Exact,2007_Lefevre_Dynamics,2024_Saha_Large,2025_Mukherjee_Hydrodynamics} and obtain the transition probability determining the microscopic evolution of the system. It is given by the sum of probabilities of all possible histories between $\boldsymbol{n}_\text{ini}$ and $\boldsymbol{n}_\text{fin}$ which we formally write as a constrained path integral
\begin{equation}
\Pr\left(\boldsymbol{n}_\text{fin}\middle|\boldsymbol{n}_\text{ini}\right)=\int_{\boldsymbol{n}_\text{ini}}^{\boldsymbol{n}_\text{fin}}\left[\mathcal{D}n\right]\Bigg<\prod_{k=0}^{M-1}\prod_{i=-\infty}^\infty\delta_{n_i(k\,\mathrm{d}\tau+\mathrm{d}\tau)-n_i(k\,\mathrm{d}\tau)\,,\,Y_{i-1}(k\,\mathrm{d}\tau)-Y_i(k\,\mathrm{d}\tau)}\Bigg>_{Y} \,,
\end{equation}
where $\delta_{a,b}$ is the Kronecker delta function, $M\mathrm{d}\tau=\mathcal{T}$ with an infinitesimal $\mathrm{d}\tau$, and the path integral measure
\begin{equation*}
\int\left[\mathcal{D}n\right]\equiv\prod_{k=1}^{M-1}\prod_{i=-\infty}^\infty\sum_{n_i(k\,\mathrm{d}\tau)=0}^1 
\end{equation*}
The angular bracket $\left<\dots\right>_Y$ denotes an average over all hopping events $\{Y_i(\tau)\}$ in the duration $\mathcal{T}$. Using an integral representation $\delta_{a,b}=\left(2\pi\mathrm{i}\right)^{-1}\int_{-\pi\mathrm{i}}^{\pi\mathrm{i}}\mathrm{d}z\,\mathrm{e}^{-z\left(a-b\right)}$ for integers $a$ and $b$, we write
\begin{align}
\Pr\left(\boldsymbol{n}_\text{fin}\middle|\boldsymbol{n}_\text{ini}\right)=\int_{\boldsymbol{n}_\text{ini}}^{\boldsymbol{n}_\text{fin}}\left[\mathcal{D}\widehat{n}\right]\left[\mathcal{D}n\right]&\;\mathrm{e}^{-\sum_{k=0}^{M-1}\sum_{i=-\infty}^\infty\widehat{n}_i(k\,\mathrm{d}\tau)\,\left[n_i(k\,\mathrm{d}\tau+\mathrm{d}\tau)-n_i(k\,\mathrm{d}\tau)\right]}\nonumber\\
&\Bigg<\mathrm{e}^{\sum_{k=0}^{M-1}\sum_{i=-\infty}^\infty\widehat{n}_i(k\,\mathrm{d}\tau)\,\left[Y_{i-1}(k\,\mathrm{d}\tau)-Y_i(k\,\mathrm{d}\tau)\right]}\Bigg>_{Y} \,,\label{trans_prob_in_parts}
\end{align}
with the path integral measure on $\widehat{n}$ defined as
\begin{equation*}
\int\left[\mathcal{D}\widehat{n}\right]\equiv\prod_{k=0}^{M-1}\prod_{i=-\infty}^{\infty}\frac{1}{2\pi\mathrm{i}}\int_{-\pi\mathrm{i}}^{\pi\mathrm{i}}\mathrm{d}\widehat{n}_i(k\,\mathrm{d}\tau)
\end{equation*}

For a given set of occupation variables, the average over $Y$ can be exactly computed using the probability of the hopping variables in \eqref{eq:bulk_hops}. To proceed, we use a summation by parts and write for each time-index $k$
\begin{equation}
\sum_{i=-\infty}^\infty\widehat{n}_i\,\big(Y_{i-1}-Y_i\big)=\sum_{i=-\infty}^\infty\big(\widehat{n}_{i+1}-\widehat{n}_i\big)\,Y_i\,.
\end{equation}
Using \eqref{eq:bulk_hops}, we then obtain
\begin{align}
\Big<\mathrm{e}^{(\widehat{n}_{i+1}-\widehat{n}_i)\,Y_i}\Big>_{Y_i}\simeq\mathrm{e}^{\mathrm{d}\tau\left[\left(\mathrm{e}^{\widehat{n}_{i+1}-\widehat{n}_i}-1\right)n_i\left(1-n_{i+1}\right)\left(1-n_{i+2}\right)+\left(\mathrm{e}^{-\widehat{n}_{i+1}+\widehat{n}_i}-1\right)n_{i+1}\left(1-n_i\right)\left(1-n_{i-1}\right)\right]} \,, \label{bulk_micro_derive}
\end{align}
where $\simeq$ denotes the leading-order term in the $\mathrm{d}\tau\to0$ limit. Using this result for averages in the path integral \eqref{trans_prob_in_parts}, we write in the $\mathrm{d}\tau\to0$ limit,
\begin{subequations}
\begin{equation}\label{eq:micro_Action}
\Pr\left(\boldsymbol{n}_\text{fin}\middle|\boldsymbol{n}_\text{ini}\right)=\int_{\boldsymbol{n}_\text{ini}}^{\boldsymbol{n}_\text{fin}}\left[\mathcal{D}\widehat{n}\right]\left[\mathcal{D}n\right]\mathrm{e}^{\mathcal{K}+\mathcal{H}} \,,
\end{equation}
where
\begin{align}
\mathcal{K}&=-\sum_{i=-\infty}^\infty\int_0^\mathcal{T}\mathrm{d}\tau\,\,\widehat{n}_i(\tau)\frac{\mathrm{d}n_i(\tau)}{\mathrm{d}\tau}\,,\text{ and} \label{eq:micro_kinetic}\\
\mathcal{H}&=\sum_{i=-\infty}^\infty\int_0^\mathcal{T}\mathrm{d}\tau\,\Big[\big(\mathrm{e}^{\widehat{n}_{i+1}(\tau)-\widehat{n}_i(\tau)}-1\big)\,n_i(\tau)\,\big(1-n_{i+1}(\tau)\big)\,\big(1-n_{i+2}(\tau)\big)\nonumber\\
&\qquad\qquad\qquad\quad+\big(\mathrm{e}^{-\widehat{n}_{i+1}(\tau)+\widehat{n}_i(\tau)}-1\big)\,n_{i+1}(\tau)\,\big(1-n_i(\tau)\big)\,\big(1-n_{i-1}(\tau)\big)\Big]\,. \label{eq:micro_Hamiltonian}
\end{align}
\label{eq:micro_path_int}
\end{subequations}

We propose a systematic way to derive the hydrodynamic path probability \eqref{eq:trans_prob_general} of the coarse-grained density field from the microscopic propagator in (\ref{eq:micro_path_int}). Assuming local equilibrium, the steps shown below are justified in Appendix~\ref{sec:explicit} through an explicit bottom-up construction of the path integral \eqref{eq:trans_prob_general} for the coarse-grained density field, and the final results are successfully compared to numerical simulations in Section.~\ref{sec:numeric_verify} and to known results in Table \ref{tab:trans_params}. 

We denote by $\ell \gg 1$ a large hydrodynamic scale --- large compared to spatial correlation lengths in equilibrium --- and assume that over lengthscales of order $\ell$, the system is described by a smoothly-varying coarse-grained density field $\rho(x,t)$. To leading order in $\ell$, this means that
\begin{equation}
\left<n_i(\tau)\right> \simeq\rho\bigg(\frac{i}{\ell},\frac{\tau}{\ell^2}\bigg)\label{rhoi_rho} \,,
\end{equation}
where the average $\left\langle \dots \right\rangle$ now denotes a spatial or temporal average over mesoscopic scales---much larger than the correlation length or time of the microscopic degrees of freedom in the equilibrium dynamics but much smaller than the hydrodynamic scale given by $\ell$. That the coarse-grained density field varies over timescales of order $\ell^2$ is expected for diffusion-dominated dynamics like in the SSDEP. We also assume that local statistics of the occupation variables $n_i(\tau)$---when probed over length- and time-scales small compared to $\ell$ and $\ell^2$ respectively---can be approximated using the equilibrium measure at a mean density given by the local value of the coarse-grained density field $\rho(x,t)$. This is the essence of the local equilibrium hypothesis \cite{1991_Spohn_Large,1999_Kipnis_Scaling}.

To complete our construction of the fluctuating hydrodynamics, we further require that the conjugate variables $\widehat{n}_i$ can be directly approximated by a smoothly-varying response field as
\begin{equation}
\widehat{n}_i(\tau)\simeq\widehat{\rho}\bigg(\frac{i}{\ell},\frac{\tau}{\ell^2}\bigg)\,,\label{eq:nhat rhohat}
\end{equation}
so that $\widehat{n}_{i+1}(\tau)-\widehat{n}_i(\tau) \sim \ell^{-1}$ and a gradient expansion of terms involving $\widehat{n}_i$ in (\ref{eq:micro_path_int}) can be performed. It is important to stress here that a naive expansion of the microscopic occupation numbers in terms of gradients of $\rho(x,t)$ would lead to incorrect results due to the existence of correlations in the local equilibrium measure. 

To proceed further, we expand the effective Hamiltonian (\ref{eq:micro_Hamiltonian}) of the microscopic action up to second order in the gradients of the response field. For notational convenience, we define 
\begin{subequations}\label{eq:right_left_rate_defn}
\begin{align}
r_i^+(\tau)&=n_i(\tau)\,\big(1-n_{i+1}(\tau)\big)\,\big(1-n_{i+2}(\tau)\big) \,, \\
r_i^-(\tau)&=n_{i+1}(\tau)\,\big(1-n_{i}(\tau)\big)\,\big(1-n_{i-1}(\tau)\big) \,,
\end{align}
\end{subequations}
the rates at which a particle jumps from site $i$ to $i+1$ and from site $i+1$ to $i$ respectively. We get 
\begin{align}
\mathcal{H}\simeq\sum_{i=-\infty}^{\infty}\int_{0}^{\mathcal{T}}\mathrm{d}\tau\,\Bigg[&\big(\widehat{n}_{i+1}(\tau)-\widehat{n}_{i}(\tau)\big)\big(r_i^+(\tau)-r_i^-(\tau)\big)\nonumber\\
&+\frac{\big(\widehat{n}_{i+1}(\tau)-\widehat{n}_{i}(\tau)\big)^2}{2}\big(r_i^+(\tau)+r_i^-(\tau)\big)\Bigg]\,.\label{eq:micro_Hamiltonian_expand}
\end{align}
The first term in (\ref{eq:micro_Hamiltonian_expand}) can now be expressed in terms of the discrete Laplacian of the response field, making both terms second order in the gradient expansion. In fact, the average current between sites $i$ and $i+1$ is a discrete gradient as seen from $r_i^+(\tau)-r_i^-(\tau)=h_i(\boldsymbol{n}(\tau))-h_{i+1}(\boldsymbol{n}(\tau))$ with 
\begin{equation}
h_i(\boldsymbol{n})=n_i+n_{i-1}n_{i+1}-n_{i-1}n_in_{i+1}\,.
\end{equation}
This allows us to perform an integration by part in (\ref{eq:micro_Hamiltonian_expand}) and to obtain 
\begin{align}
\mathcal{H}\simeq\sum_{i=-\infty}^\infty\int_0^\mathcal{T}\mathrm{d}\tau\,\Bigg[&\big(\widehat{n}_{i+1}(\tau)-2\widehat{n}_{i}(\tau)+\widehat{n}_{i-1}(\tau)\big)h_i(\boldsymbol{n}(\tau))\nonumber\\
&+\frac{\big(\widehat{n}_{i+1}(\tau)-\widehat{n}_{i}(\tau)\big)^2}{2}\big(r_i^+(\tau)+r_i^-(\tau)\big)\Bigg]\,. \label{eq:micro_Hamiltonian_expand_bis}
\end{align}
The action in (\ref{eq:micro_path_int}) can now be straightforwardly expressed in terms of the smoothly-varying response field $\widehat{\rho}$ and its gradients, but the introduction of the coarse-grained density field requires more care. In (\ref{eq:micro_Hamiltonian_expand_bis}), both terms inside the sum scale as $\ell^{-2}$. Furthermore, the contributions coming from the response field are slowly varying over scales of order $\ell$. The law of large numbers and the local equilibrium hypothesis, see Appendix~\ref{sec:explicit}, thus allow us to obtain the fluctuating hydrodynamics action to leading order in $\ell$ by replacing the contributions coming from the microscopic occupancy numbers by their average over the local equilibrium measure. In terms of the rescaled space and time coordinates $(i,\tau)\to(x,t)=\left(i/\ell,\tau/\ell^2\right)$, we thus get the transition probability of the coarse-grained density field as
\begin{subequations}
\begin{equation}\label{eq:ssdep_path_int}
\Pr\left(\rho(x,T)\middle|\rho(x,0)\right)\simeq\int_{\rho(x,0)}^{\rho(x,T)}\left[\mathcal{D}\widehat{\rho}\right]\left[\mathcal{D}\rho\right]\mathrm{e}^{-\ell \mathcal{S}[\rho,\widehat{\rho}]} \,,
\end{equation}
with $T=\mathcal{T}/\ell^2$ and where the action is expressed from (\ref{eq:micro_kinetic}) and (\ref{eq:micro_Hamiltonian_expand_bis}) as
\begin{align}\label{eq:ssdep_action}
\mathcal{S}[\rho,\widehat{\rho}]=\int_0^T\mathrm{d}t\int_{-\infty}^\infty\mathrm{d}x\,\Bigg[&\widehat{\rho}(x,t)\partial_t\!\left<n\right>_{x,t}+\partial_x\widehat{\rho}(x,t)\partial_x\!\left<h\right>_{x,t}\nonumber\\
&-\frac{\left(\partial_x\widehat{\rho}(x,t)\right)^2}{2}\Big(\left<r^+\right>_{x,t} + \left<r^-\right>_{x,t}\Big)\Bigg]\,. 
\end{align}
\end{subequations}
Here, the measure $\langle\dots\rangle_{x,t}$ corresponds to equilibrium averages at a mean density $\rho(x,t)$. Unlike in the SSEP, the two-point correlation does not factorize \cite{2004_Schonherr_Exclusion,2013_Krapivsky_Dynamics} and the averaging over the local equilibrium measure is more involved than simply replacing the occupation numbers by the coarse-grained density field. In fact \cite{2004_Schonherr_Exclusion,2013_Krapivsky_Dynamics}, 
\begin{equation}
\left<n_i n_{i+2}\right>_{x,t}=\frac{\rho(x,t)^2}{1-\rho(x,t)}\text{ and }\left<n_{i}n_{i+1}n_{i+2}\right>_{x,t}= 0 \,.
\end{equation}
To conclude, we thus get
\begin{equation}\label{eq:ssdep_action_2}
\mathcal{S}[\rho,\widehat{\rho}]=\int_0^T\mathrm{d}t\int_{-\infty}^\infty\mathrm{d}x\,\Bigg[\widehat{\rho}\partial_t\rho+\frac{1}{(1-\rho)^2}\partial_x\widehat{\rho}\partial_x\rho-\frac{\rho(1-2\rho)}{1-\rho}(\partial_x \widehat{\rho})^2\Bigg]\,.
\end{equation}
The corresponding fluctuating hydrodynamic equation \eqref{eq:fhd_general} is then explicitly given as
\begin{equation}\label{eq:ssdep_fhd}
\partial_t\rho=\partial_x\left[\frac{1}{(1-\rho)^2}\partial_x\rho\right]+\frac{1}{\sqrt{\ell}}\partial_x\left[\sqrt{\frac{2\rho(1-2\rho)}{1-\rho}}\eta\right] \,,
\end{equation}
where $\eta$ is a Gaussian noise with mean $\left<\eta(x,t)\right>=0$ and covariance $\left<\eta(x,t)\eta(x',t')\right>=\delta(x-x')\delta(t-t')$. The expression of $D(\rho)$ and $\sigma(\rho)$ from \eqref{eq:ssdep_fhd} agrees with reported values in \cite{2004_Schonherr_Exclusion,2013_Krapivsky_Dynamics,2017_Baek_Dynamical,2023_Rizkallah_Duality}.

\subsection{Generalization to Symmetric Simple Multiple Exclusion Process}

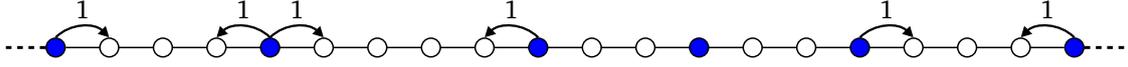
\begin{figure}
\centering
\resizebox{\linewidth}{!}
{
\begin{tikzpicture}
\node [circle, draw=black, thick, fill=blue, inner sep=0pt, minimum size=10pt] (0) at (0,0) {};
\node [circle, draw=black, thick, fill=blue, inner sep=0pt, minimum size=10pt] (19) at (19,0) {};
\draw[thick] (0) -- (19);
\node [circle, draw=black, thick, fill=white, inner sep=0pt, minimum size=10pt] (1) at (1,0) {};
\node [circle, draw=black, thick, fill=white, inner sep=0pt, minimum size=10pt] (2) at (2,0) {};
\node [circle, draw=black, thick, fill=white, inner sep=0pt, minimum size=10pt] (3) at (3,0) {};
\node [circle, draw=black, thick, fill=blue, inner sep=0pt, minimum size=10pt] (4) at (4,0) {};
\node [circle, draw=black, thick, fill=white, inner sep=0pt, minimum size=10pt] (5) at (5,0) {};
\node [circle, draw=black, thick, fill=white, inner sep=0pt, minimum size=10pt] (6) at (6,0) {};
\node [circle, draw=black, thick, fill=white, inner sep=0pt, minimum size=10pt] (7) at (7,0) {};
\node [circle, draw=black, thick, fill=white, inner sep=0pt, minimum size=10pt] (8) at (8,0) {};
\node [circle, draw=black, thick, fill=blue, inner sep=0pt, minimum size=10pt] (9) at (9,0) {};
\node [circle, draw=black, thick, fill=white, inner sep=0pt, minimum size=10pt] (10) at (10,0) {};
\node [circle, draw=black, thick, fill=white, inner sep=0pt, minimum size=10pt] (11) at (11,0) {};
\node [circle, draw=black, thick, fill=blue, inner sep=0pt, minimum size=10pt] (12) at (12,0) {};
\node [circle, draw=black, thick, fill=white, inner sep=0pt, minimum size=10pt] (13) at (13,0) {};
\node [circle, draw=black, thick, fill=white, inner sep=0pt, minimum size=10pt] (14) at (14,0) {};
\node [circle, draw=black, thick, fill=blue, inner sep=0pt, minimum size=10pt] (15) at (15,0) {};
\node [circle, draw=black, thick, fill=white, inner sep=0pt, minimum size=10pt] (16) at (16,0) {};
\node [circle, draw=black, thick, fill=white, inner sep=0pt, minimum size=10pt] (17) at (17,0) {};
\node [circle, draw=black, thick, fill=white, inner sep=0pt, minimum size=10pt] (18) at (18,0) {};
\draw [-{Triangle[width=2.0mm, length=2.0mm]}, very thick] (0.north) to [out=45,in=135] node [midway, above]{\Large$1$} (1.north);
\draw [-{Triangle[width=2.0mm, length=2.0mm]}, very thick] (4.north) to [out=135,in=45] node [midway, above]{\Large$1$} (3.north);
\draw [-{Triangle[width=2.0mm, length=2.0mm]}, very thick] (4.north) to [out=45,in=135] node [midway, above]{\Large$1$} (5.north);
\draw [-{Triangle[width=2.0mm, length=2.0mm]}, very thick] (9.north) to [out=135,in=45] node [midway, above]{\Large$1$} (8.north);
\draw [-{Triangle[width=2.0mm, length=2.0mm]}, very thick] (15.north) to [out=45,in=135] node [midway, above]{\Large$1$} (16.north);
\draw [-{Triangle[width=2.0mm, length=2.0mm]}, very thick] (19.north) to [out=135,in=45] node [midway, above]{\Large$1$} (18.north);
\draw[ultra thick, dashed] (0) -- (-1,0);
\draw[ultra thick, dashed] (19) -- (20,0);
\end{tikzpicture}
}
\caption{\textbf{SSTEP:} A schematic of the dynamics of the symmetric simple \emph{triple} exclusion process ($M=3$) on a one-dimensional lattice. Filled circles denote occupied sites, while open circles represent empty sites. Arrows indicate the allowed particle hops, each occurring with rate~$1$.}
\label{fig:sstep_scheme}
\end{figure}

Next, we generalize our derivation of fluctuating hydrodynamics to a scenario where the volume exclusion extends to multiple neighboring sites \cite{1968_Macdonald_Kinetics,1969_Macdonald_Concerning,2004_Schonherr_Exclusion,2013_Krapivsky_Dynamics} around the site occupied by a particle. In this model, a particle simultaneously occupies $(M-1)$ adjacent sites on either side of its current site. We refer to this generalization as the Symmetric Simple Multiple Exclusion Process (SSMEP). The system evolves in continuous time respecting the extended hard-core constraint where a particle at the $i$-th site hops to the $(i+1)$-th site provided that the sites $i+1,\,i+2,\,\dots,\,i+M$ are vacant. Similarly, the particle at $i$-th site hops to the $(i-1)$-th site provided that the sites $i-1,\,i-2,\,\dots,\,i-M$ are all vacant. A schematic of the case for $M=3$ (the Symmetric Simple Triple Exclusion Process) is shown in Fig. \ref{fig:sstep_scheme}.

Following the steps outlined previously for the SSDEP, we arrive at the coarse-grained transition probability and the corresponding fluctuating hydrodynamic equation, respectively
\begin{equation}\label{eq:ssmep_path_int}
\Pr\left(\rho(x,T)\middle|\rho(x,0)\right)\simeq\int_{\rho(x,0)}^{\rho(x,T)}\left[\mathcal{D}\widehat{\rho}\right]\left[\mathcal{D}\rho\right]\mathrm{e}^{-\ell\int_0^T\mathrm{d}t\int_{-\infty}^\infty\mathrm{d}x\left[\widehat{\rho}\partial_t\rho+\frac{1}{[1-(M-1)\rho]^2}\partial_x\widehat{\rho}\,\partial_x\rho-\frac{\rho(1-M\rho)}{1-(M-1)\rho}(\partial_x\widehat{\rho})^2\right]} \,,
\end{equation}
and
\begin{equation}\label{eq:ssmep_fhd}
\partial_t\rho=\partial_x\left\{\frac{1}{[1-(M-1)\rho]^2}\partial_x\rho\right\}+\frac{1}{\sqrt{\ell}}\partial_x\left[\sqrt{\frac{2\rho(1-M\rho)}{1-(M-1)\rho}}\eta\right] \,.
\end{equation}
The transport coefficients for the SSMEP thus read,
\begin{equation}\label{eq:ssmep_trans_param}
D(\rho)=\frac{1}{[1-(M-1)\rho]^2}\text{ and }\sigma(\rho)=\frac{2\rho(1-M\rho)}{1-(M-1)\rho} \,.
\end{equation}
Substituting $M=1$, we recover the results for the SSEP, corresponding to the case with no exclusion beyond the site being currently occupied, while $M=2$ leads to our results for the SSDEP.

Let us trace out the crucial steps in the coarse-graining for the Symmetric Simple Triple Exclusion Process (SSTEP) which corresponds to the $M=3$ case. For the SSTEP, the right and left hop rates across the $(i,i+1)$-bond in \eqref{eq:right_left_rate_defn} modifies to
\begin{subequations}\label{eq:sstep_rgt_lft_rate}
\begin{align}
r_i^+&=n_i(1-n_{i+1})(1-n_{i+2})(1-n_{i+3})\\
r_i^-&=n_{i+1}(1-n_i)(1-n_{i-1})(1-n_{i-2})
\end{align}
This reveals a gradient structure where the average current across the $(i,i+1)$-bond is a discrete gradient as seen by writing $r_i^+-r_i^-=h_i(\boldsymbol{n})-h_{i+1}(\boldsymbol{n})$ with
\begin{align}\label{eq:sstep_loc_h_fn}
h_i(\boldsymbol{n})=&\,n_i+n_{i-2}n_{i+1}+n_{i-1}n_{i+2}+n_{i-1}n_{i+1}-n_{i-2}n_{i-1}n_{i+1}-n_{i-1}n_{i}n_{i+2}\nonumber\\
&-n_{i-2}n_{i}n_{i+1}-n_{i-1}n_{i+1}n_{i+2}-n_{i-1}n_{i}n_{i+1}+n_{i-2}n_{i-1}n_{i}n_{i+1}+n_{i-1}n_{i}n_{i+1}n_{i+2}.
\end{align}
\end{subequations}
This gradient structure allows us to write the hydrodynamic action for the SSTEP in a similar form in \eqref{eq:ssdep_action} with $h$ and $r^\pm$ given in \eqref{eq:sstep_loc_h_fn} and \eqref{eq:sstep_rgt_lft_rate}, respectively. For computing the local equilibrium averages in \eqref{eq:ssdep_action} we use the known local equilibrium state \cite{2004_Schonherr_Exclusion,2013_Krapivsky_Dynamics} for SSTEP and write
\begin{subequations}
\begin{equation}
\left<h_i\right>_{x,t}=\left<n_i\right>_{x,t}+\left<n_{i-2}n_{i+1}\right>_{x,t}+\left<n_{i-1}n_{i+2}\right>_{x,t}=\frac{\rho(x,t)}{1-2\rho(x,t)}
\end{equation}
and
\begin{align}
\left<r_i^+\right>_{x,t}&=\left<n_i\right>_{x,t}-\left<n_in_{i+3}\right>_{x,t}=\frac{\rho(x,t)\left(1-3\rho(x,t)\right)}{1-2\rho(x,t)}\\
\left<r_i^-\right>_{x,t}&=\left<n_{i+1}\right>_{x,t}-\left<n_{i-2}n_{i+1}\right>_{x,t}=\frac{\rho(x,t)\left(1-3\rho(x,t)\right)}{1-2\rho(x,t)}
\end{align}
\end{subequations}
where we used the fact that the rest of the terms in \eqref{eq:sstep_rgt_lft_rate} and \eqref{eq:sstep_loc_h_fn} contribute zero owing to the $3$-site exclusion. The resulting action \eqref{eq:ssmep_path_int} with $M=3$ gives the associated fluctuating hydrodynamics for SSTEP.

For SSMEP the crucial step in averaging the microscopic occupation variables with the local equilibrium measure comes from the fact that the only non-vanishing contribution comes from terms containing the product of occupation variables of sites separated by exactly $(M-1)$ sites and are given by \cite{2013_Krapivsky_Dynamics,2004_Schonherr_Exclusion}
\begin{align}
\left<n_i \, n_{i+M}\right>_{x,t}&=\frac{\rho^2(x,t)}{1-(M-1)\rho(x,t)} \,.
\end{align}
The denominator accounts for the volume exclusion where the sites $i$ and $i+M$ can be simultaneously occupied if and only if all $(M-1)$ intermediate sites are vacant. The rest of the calculation for arriving at \eqref{eq:ssmep_fhd} is straightforward.

We now make an important observation. Let us take the lattice spacing to be of unit length. Then, the SSMEP describes particles of effective length $M\equiv a$ that hop in discrete steps of unit size. Taking the limit of $M\gg1$, we find that the transport coefficients in \eqref{eq:ssmep_trans_param} reduce to
\begin{equation}
D(\rho)\simeq \frac{1}{(1-M\rho)^2}\;\text{and}\;\sigma(\rho)\simeq 2\rho \,,
\end{equation}
which coincide with those of a gas of Brownian hard rods \cite{2005_Schonherr_Hard,2005_Lin_From,2023_Rizkallah_Duality} (see Table \ref{tab:trans_params}). This confirms the intuition that a suitable continuous limit (large-$M$ limit) of the lattice-model of extended hard-core particles, describes the large-scale dynamics of the continuum model of Brownian hard rods of finite length.

\subsection{Symmetric Simple Partial Exclusion Process}

We next consider a variation \cite{1994_Schutz_Non,2008_Tailleur_Mapping,2016_Baek_Extreme,2023_Franceschini_Hydrodynamical} of the symmetric simple exclusion process, where the hard-core constraint on at most one particle per site at a given moment is relaxed such that each site can accommodate as many as $N$ particles. Particles hop to either of its nearest-neighboring sites at a rate which is proportional to the number of vacancies in the target site \cite{1994_Schutz_Non,2008_Tailleur_Mapping,2016_Baek_Extreme,2023_Franceschini_Hydrodynamical}. More precisely, a hopping-event occurs from site $i$ to site $i\pm1$ at a rate $n_i(N-n_{i\pm1})$, i.e.,
\begin{subequations}\label{eq:sspep_rgt_lft_rate}
\begin{align}
r_i^+&=n_i(N-n_{i+1})\\
r_i^-&=n_{i+1}(N-n_i)
\end{align}
We refer to this dynamics as the Symmetric Simple Partial Exclusion Process (SSPEP).

Similar to the SSDEP, for this model as well, the average current across the bond $(i,i+1)$ exhibits a discrete gradient structure with the local $h$-function,
\begin{equation}
h_i(\boldsymbol{n})=Nn_i\,.
\end{equation}
\end{subequations}
The coarse-graining is similarly done using the path-integral approach as for the SSDEP in Section.~\ref{sec:ssdep} with the changes coming only at the equilibrium averages of the hopping rates. For the SSPEP, we use the known \cite{2013_Carinci_Duality,2022_Floreani_Orthogonal} local equilibrium measure to compute
\begin{subequations}
\begin{equation}
\left<h_i(\boldsymbol{n})\right>_{x,t}=N\!\left<n_i\right>_{x,t}=N\rho(x,t)
\end{equation}
and
\begin{align}
\left<r^+\right>_{x,t}&=N\!\left<n_i\right>_{x,t}-\left<n_in_{i+1}\right>_{x,t}=\rho(x,t)\left(N-\rho(x,t)\right)\,,\\
\left<r^-\right>_{x,t}&=N\!\left<n_{i+1}\right>_{x,t}-\left<n_{i+1}n_i\right>_{x,t}=\rho(x,t)\left(N-\rho(x,t)\right).
\end{align}
\end{subequations}
This leads to the hydrodynamic action \eqref{eq:hydro_hamiltonian} and subsequently, the fluctuating hydrodynamic equation \eqref{eq:fhd_general} with the transport coefficients $D(\rho)=N$ and $\sigma(\rho)=2\rho(N-\rho)$ (see Table \ref{tab:trans_params}).

Another variation of this model \cite{2014_Arita_Generalized} is to set the hopping rate of a particle to $1$ irrespective of the number of vacancies in the target site, which amounts to the rate of hopping from site $i$ to sites $i\pm1$ being $n_i\left\{1-\left[n_{i\pm1}(n_{i\pm1}-1)\dots(n_{i\pm1}-N+1)\right]/N!\right\}$ with exclusion constraint incorporated. This choice breaks the gradient structure of the hopping rates \cite{2014_Arita_Generalized,2017_Arita_Variational} which is essential in our coarse-graining method.

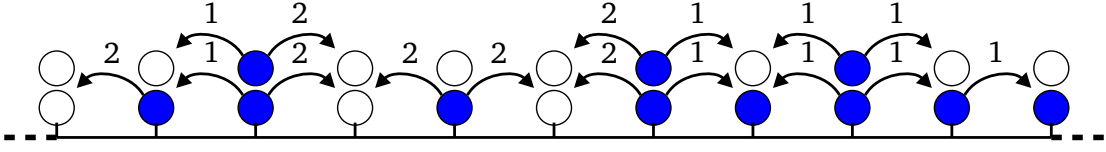
\begin{figure}
\centering
\resizebox{\linewidth}{!}
{
\begin{tikzpicture}
\node [circle, draw=black, thin, fill=white, inner sep=0pt, minimum size=10pt] (0) at (0,0) {};
\node [circle, draw=black, thin, fill=white, inner sep=0pt, minimum size=10pt] (0a) at (0,0.4) {};
\node [circle, draw=black, thin, fill=blue, inner sep=0pt, minimum size=10pt] (10) at (10,0) {};
\node [circle, draw=black, thin, fill=white, inner sep=0pt, minimum size=10pt] (10a) at (10,0.4) {};
\node [circle, draw=black, thin, fill=blue, inner sep=0pt, minimum size=10pt] (1) at (1,0) {};
\node [circle, draw=black, thin, fill=white, inner sep=0pt, minimum size=10pt] (1a) at (1,0.4) {};
\node [circle, draw=black, thin, fill=blue, inner sep=0pt, minimum size=10pt] (2) at (2,0) {};
\node [circle, draw=black, thin, fill=blue, inner sep=0pt, minimum size=10pt] (2a) at (2,0.4) {};
\node [circle, draw=black, thin, fill=white, inner sep=0pt, minimum size=10pt] (3) at (3,0) {};
\node [circle, draw=black, thin, fill=white, inner sep=0pt, minimum size=10pt] (3a) at (3,0.4) {};
\node [circle, draw=black, thin, fill=blue, inner sep=0pt, minimum size=10pt] (4) at (4,0) {};
\node [circle, draw=black, thin, fill=white, inner sep=0pt, minimum size=10pt] (4a) at (4,0.4) {};
\node [circle, draw=black, thin, fill=white, inner sep=0pt, minimum size=10pt] (5) at (5,0) {};
\node [circle, draw=black, thin, fill=white, inner sep=0pt, minimum size=10pt] (5a) at (5,0.4) {};
\node [circle, draw=black, thin, fill=blue, inner sep=0pt, minimum size=10pt] (6) at (6,0) {};
\node [circle, draw=black, thin, fill=blue, inner sep=0pt, minimum size=10pt] (6a) at (6,0.4) {};
\node [circle, draw=black, thin, fill=blue, inner sep=0pt, minimum size=10pt] (7) at (7,0) {};
\node [circle, draw=black, thin, fill=white, inner sep=0pt, minimum size=10pt] (7a) at (7,0.4) {};
\node [circle, draw=black, thin, fill=blue, inner sep=0pt, minimum size=10pt] (8) at (8,0) {};
\node [circle, draw=black, thin, fill=blue, inner sep=0pt, minimum size=10pt] (8a) at (8,0.4) {};
\node [circle, draw=black, thin, fill=blue, inner sep=0pt, minimum size=10pt] (9) at (9,0) {};
\node [circle, draw=black, thin, fill=white, inner sep=0pt, minimum size=10pt] (9a) at (9,0.4) {};
\draw[thick] (0,-0.3) -- (10,-0.3);
\foreach \x in {0,1,...,10} {\draw[thick] (\x,-0.3) -- (\x,-0.15);}
\draw[ultra thick, dashed] (0,-0.3) -- (-0.6,-0.3);
\draw[ultra thick, dashed] (10,-0.3) -- (10.6,-0.3);
\draw [-{Triangle[width=1.5mm, length=1.5mm]}, thick] (1.north west) to [out=120,in=45] node [midway, above]{\footnotesize$2$} (0.2,0.2);
\draw [-{Triangle[width=1.5mm, length=1.5mm]}, thick] (2.north west) to [out=120,in=45] node [midway, above]{\footnotesize$1$} (1.2,0.2);
\draw [-{Triangle[width=1.5mm, length=1.5mm]}, thick] (2a.north west) to [out=120,in=45] node [midway, above]{\footnotesize$1$} (1.2,0.6);
\draw [-{Triangle[width=1.5mm, length=1.5mm]}, thick] (2.north east) to [out=60,in=135] node [midway, above]{\footnotesize$2$} (2.8,0.2);
\draw [-{Triangle[width=1.5mm, length=1.5mm]}, thick] (2a.north east) to [out=60,in=135] node [midway, above]{\footnotesize$2$} (2.8,0.6);
\draw [-{Triangle[width=1.5mm, length=1.5mm]}, thick] (4.north west) to [out=120,in=45] node [midway, above]{\footnotesize$2$} (3.2,0.2);
\draw [-{Triangle[width=1.5mm, length=1.5mm]}, thick] (4.north east) to [out=60,in=135] node [midway, above]{\footnotesize$2$} (4.8,0.2);
\draw [-{Triangle[width=1.5mm, length=1.5mm]}, thick] (6.north west) to [out=120,in=45] node [midway, above]{\footnotesize$2$} (5.2,0.2);
\draw [-{Triangle[width=1.5mm, length=1.5mm]}, thick] (6a.north west) to [out=120,in=45] node [midway, above]{\footnotesize$2$} (5.2,0.6);
\draw [-{Triangle[width=1.5mm, length=1.5mm]}, thick] (6.north east) to [out=60,in=135] node [midway, above]{\footnotesize$1$} (6.8,0.2);
\draw [-{Triangle[width=1.5mm, length=1.5mm]}, thick] (6a.north east) to [out=60,in=135] node [midway, above]{\footnotesize$1$} (6.8,0.6);
\draw [-{Triangle[width=1.5mm, length=1.5mm]}, thick] (8.north west) to [out=120,in=45] node [midway, above]{\footnotesize$1$} (7.2,0.2);
\draw [-{Triangle[width=1.5mm, length=1.5mm]}, thick] (8a.north west) to [out=120,in=45] node [midway, above]{\footnotesize$1$} (7.2,0.6);
\draw [-{Triangle[width=1.5mm, length=1.5mm]}, thick] (8.north east) to [out=60,in=135] node [midway, above]{\footnotesize$1$} (8.8,0.2);
\draw [-{Triangle[width=1.5mm, length=1.5mm]}, thick] (8a.north east) to [out=60,in=135] node [midway, above]{\footnotesize$1$} (8.8,0.6);
\draw [-{Triangle[width=1.5mm, length=1.5mm]}, thick] (9.north east) to [out=60,in=135] node [midway, above]{\footnotesize$1$} (9.8,0.2);
\end{tikzpicture}
}
\caption{\textbf{SSPEP}: A schematic of the dynamics of the Symmetric Simple Partial Exclusion Process for the case where threshold occupation per site is $N=2$. Particles are indicated by filled circles, while the empty circles denote vacancies. Allowed nearest neighbor hops are indicated by arrows with the corresponding hopping rates for individual particles.}
\label{fig:sshep_scheme}
\end{figure}

\section{Langevin dynamics: Interacting Brownian particles} \label{sec:continuum}

We now consider a system of $N$ Brownian particles moving on the real line and interacting through a symmetric pairwise potential $V(x-y)$. For our discussion we consider short-range potentials \cite{1982_Frohlich_The,1989_Ruelle_Statistical,2004_Cuesta_General} where $V(x)$ is decaying faster than $1/x^2$ therefore avoiding equilibrium phase transitions. The stochastic dynamics of the $i$-th particle in the overdamped limit is given by the Langevin equation
\begin{equation}\label{eq:Langevian_eq}
\dot{X}_i(\tau)=-\mu\sum_{j\ne i}\partial_{X_i}V\left(X_i(\tau)-X_j(\tau)\right)+\sqrt{2D_0}\zeta_i(\tau) \,,
\end{equation}
where $\mu$ denotes the particle mobility and $D_0$ is the bare diffusion coefficient, related to $\mu$ via the Einstein–Smoluchowski relation $D_0=\mu/\beta$ at thermal equilibrium. The noise terms $\zeta_i(\tau)$ are independent Gaussian white noises with zero mean $\left<\zeta_i(\tau)\right>=0$ and unit covariance $\left<\zeta_i(\tau)\zeta_{i'}(\tau')\right>=\delta_{i,i'}\delta(\tau-\tau')$.

The Dean-Kawasaki equation \cite{1994_Kawasaki_Stochastic,1996_Dean_Langevin,2025_Illien_the} provides an exact stochastic evolution for the microscopic empirical density
\begin{equation}
\varrho(X,\tau)\equiv\sum_{i=1}^N\delta(X-X_i(\tau)) \,.
\end{equation}
\begin{subequations}
\label{eq:DK full}
It takes the form of a continuity equation
\begin{equation}
\partial_\tau\varrho(X,\tau)=-\partial_Xj(X,\tau) \,,
\end{equation}
where the current is decomposed into a deterministic component, describing average transport, and a fluctuating component, capturing the intrinsic noise from the underlying Brownian dynamics, $j=j_{\text{det}}+j_{\text{fluc}}$. The deterministic contribution comes from diffusive flux due to the density gradient, and an interaction-driven drift arising from pairwise forces
\begin{equation}\label{eq:det_micro_curr}
j_{\text{det}}(X,\tau)=-D_0\partial_X\varrho(X,\tau)-\mu\varrho(X,\tau)\int_{-\infty}^\infty\mathrm{d}Y\partial_XV(X-Y)\varrho(Y,\tau) \,,
\end{equation}
and the fluctuating component
\begin{equation}
j_{\text{fluc}}(X,\tau)=\sqrt{2D_0\varrho(X,\tau)}\xi(X,\tau) \,,
\end{equation}
\end{subequations}
where $\xi(X,\tau)$ is a Gaussian white noise having zero mean $\left<\xi(X,\tau)\right>=0$ and is delta-correlated in space and time with unit covariance $\left<\xi(X,\tau)\xi(X',\tau')\right>=\delta(X-X')\delta(\tau-\tau')$. In writing \eqref{eq:det_micro_curr} we follow the It\^o convention. 

A coarse-grained description of the Dean-Kawasaki equation \eqref{eq:DK full} is the fluctuating hydrodynamics equation \eqref{eq:fhd_general} with the diffusivity $D(\rho)=\beta P'(\rho)$ and linear mobility $\sigma(\rho)=2\rho$ (see Table \ref{tab:trans_params}). It is evident that under coarse-graining the interaction term in \eqref{eq:det_micro_curr} re-normalizes the bare diffusivity $D_0$ which a simple rescaling of coordinates fails to offer. In the following we show how this coarse-graining can be systematically performed leading to \eqref{eq:fhd_general}.

We follow a scheme similar to that described in Section.~\ref{sec:lattice} for the lattice model. The transition amplitude between initial $\varrho_{\text{ini}}\equiv\varrho(X,0)$ and a final $\varrho_{\text{fin}}\equiv\varrho(X,\mathcal{T})$ following Dean-Kawasaki equation \eqref{eq:DK full} has a path-integral representation \cite{2015_Krapivsky_Melting,2025_Bon_Non,1973_Martin_Statistical,1976_Janssen_On,1976_Dominicis_Techniques,1978_Dominicis_Field}
\begin{subequations}
\begin{equation}\label{eq:DK_action_micro}
\Pr\left(\varrho_{\text{fin}}\middle|\varrho_{\text{ini}}\right)=\int_{\varrho_{\text{ini}}}^{\varrho_{\text{fin}}}\left[\mathcal{D}\widehat{\varrho}\right]\left[\mathcal{D}\varrho\right]\mathrm{e}^{-\int_0^{\mathcal{T}}\mathrm{d}\tau\left[\int_{-\infty}^\infty\mathrm{d}X\left(\widehat{\varrho}\partial_\tau\varrho\right)-\mathcal{H}\left(\widehat{\varrho},\varrho\right)\right]}
\end{equation}
with an effective microscopic Hamiltonian
\begin{equation}\label{eq:micro_DK_Hamiltonian}
\mathcal{H}\left(\widehat{\varrho},\varrho\right)=\int_{-\infty}^\infty\mathrm{d}X\partial_X\widehat{\varrho}\left(D_0\varrho\partial_X\widehat{\varrho}+j_{\text{det}}\right).
\end{equation}
\end{subequations}

In overdamped interacting Brownian systems, the dynamics is diffusive and thus at large times $\tau\gg\ell^2\gg 1$, fluctuations in regions of length smaller than $\ell$ effectively reach a local equilibrium \cite{1991_Spohn_Large,1999_Kipnis_Scaling}. By construction the length $\ell$ is much larger than the equilibrium correlation length. The local statistics of $\varrho(X,\tau)$ can thus be approximated by the local equilibrium measure corresponding to a smoothly varying coarse-grained density profile
\begin{equation}\label{eq:avgvarrho_rho}
\left<\varrho(X,\tau)\right> \simeq\rho\left(\frac{X}{\ell},\frac{\tau}{D_0\ell^2}\right) \,.
\end{equation}
As in Sec.~\ref{sec:lattice}, we further assume that we can take the response field as being smoothly varying
\begin{equation}\label{eq:hatvarrho_hatrho}
\widehat{\varrho}(X,\tau)\simeq\widehat{\rho}\left(\frac{X}{\ell},\frac{\tau}{D_0\ell^2}\right) \,.
\end{equation}
In the microscopic action (\ref{eq:DK_action_micro}-\ref{eq:micro_DK_Hamiltonian}), the only nonlinear term in the density field comes from the deterministic current $j_{\text{det}}$. As for the SSDEP, we note that $j_{\text{det}}$ has a gradient structure \cite{1991_Spohn_Large} since it can be expressed as the divergence of the Irving-Kirkwood tensor \cite{1950_Irving_The}, 
\begin{align}\label{eq:Irv-Kirk}
j_{\text{det}} & = - \partial_ X \Bigg[ D_0 \varrho(X,\tau) - \frac{\mu}{2} \int_{-\infty}^\infty\mathrm{d}Y\, V'(Y)Y \int_0^1 \mathrm{d}\lambda \, \varrho(X-\lambda Y,\tau)\varrho(X+(1-\lambda)Y,\tau) \Bigg] \,.
\end{align}
The gradient structure extends to higher dimensions for isotropic pair-potential \cite{1950_Irving_The}. This structure (\ref{eq:Irv-Kirk}) makes it straightforward to follow the coarse-graining procedure outlined in Sec.~\ref{sec:lattice}.

Re-scaling space and time in \eqref{eq:DK_action_micro} as $\left(X,\tau\right)\to\left(x,t\right)=\left(X/\ell,\tau/(D_0\ell^2)\right)$, we write
the transition amplitude for the coarse-grained density field
\begin{subequations}
\begin{equation}\label{eq:brownian_path_int}
\Pr\left(\rho(x,T)\middle|\rho(x,0)\right)=\int_{\rho(x,0)}^{\rho(x,T)}\left[\mathcal{D}\widehat{\rho}\right]\left[\mathcal{D}\rho\right]\mathrm{e}^{-\ell\mathcal{S}[\rho,\widehat{\rho}]}\,,
\end{equation}
where we denoted $T=\mathcal{T}/\left(D_0\ell^2\right)$, and with the action
\begin{align}\label{eq:action_hydro_DK1}
\mathcal{S}[\rho,\widehat{\rho}]=&\int_0^T \mathrm{d}t\int_{-\infty}^\infty\mathrm{d}x\,\Bigg\{\widehat{\rho}\partial_t\rho-\partial_x^2\widehat{\rho}\bigg[\rho-\frac{\beta}{2}\int_{-\infty}^\infty\mathrm{d}Y\,\bigg(V'(Y)Y\nonumber\\
&\times\int_{0}^{1}\mathrm{d}\lambda\,\big<\varrho(X-\lambda Y,\tau)\varrho(X+(1-\lambda)Y,\tau)\big>_{x,t}\bigg)\bigg]-\big(\partial_x\widehat{\rho}\big)^2\rho\Bigg\}\,. 
\end{align}
\end{subequations}
Equation (\ref{eq:action_hydro_DK1}) features the two-point correlation function $\rho^{(2)}$ at a separation $Y$ of a homogeneous equilibrium system at density $\rho(x,t)$,
\begin{equation}
\left\langle \varrho(X-\lambda Y,\tau)\varrho(X+(1-\lambda)Y,\tau) \right\rangle_{x,t} = \rho^{(2)}[\rho(x,t)](Y) \,.
\end{equation}
We see that the hydrodynamic action (\ref{eq:action_hydro_DK1}) can be written in terms of the equilibrium equation of state for the pressure \cite{1950_Irving_The}
\begin{equation}
\beta P(\rho)=\rho-\frac{\beta}{2}\int_{-\infty}^\infty\mathrm{d}Y \, Y\, V'(Y)\rho^{(2)}[\rho]\left(Y\right) \,,
\end{equation}
and we get
\begin{align}
\mathcal{S}[\rho,\widehat{\rho}] = \int_0^T & \mathrm{d}t\int_{-\infty}^\infty\mathrm{d}x\left[\widehat{\rho}\partial_t\rho+\beta P'(\rho)\partial_x\widehat{\rho}\partial_x\rho-\rho(\partial_x\widehat{\rho})^2\right]\,. 
\end{align}
Two comments are in order. First, by comparing the transition probability of the coarse-grained profile for interacting Brownian particles \eqref{eq:brownian_path_int} with that for arbitrary diffusive systems \eqref{eq:trans_prob_general}, we establish a general relation between the pressure, a static equilibrium observable, and the diffusivity, a dynamic non-equilibrium observable as $D(\rho)=\beta P'(\rho)$. Second, we see that the collective mobility of interacting Brownian particles is given by $\sigma(\rho)=2\rho$ regardless of the specific form of the two-body interaction. These results were previously argued based on the fluctuation-dissipation relation \cite{2014_Krapivsky_Large,2015_Krapivsky_Tagged,2025_Illien_the,2025_Grabsch_Exact,Sadhu2023_ICTS_Talk}.

The fluctuating hydrodynamic equation corresponding to the hydrodynamic action in \eqref{eq:brownian_path_int} explicitly reads
\begin{equation}\label{eq:brownian_fhd}
\partial_t\rho=\partial_x\left(\beta P'(\rho)\partial_x\rho\right)+\frac{1}{\sqrt{\ell}} \partial_x\left(\sqrt{2\rho}\eta\right)\,,
\end{equation}
with $\left<\eta(x,t)\right>=0$ and $\left<\eta(x,t)\eta(x',t')\right>=\delta(x-x')\delta(t-t')$. For Brownian hard-rods of length $a$, with interaction potential
\begin{equation}\label{eq:HardRod_pot}
V(r)=
\begin{cases}
\infty & \text{for }\left|r\right|<a \,, \\[6pt]
0 & \text{for }\left|r\right|\geq a \,,
\end{cases}
\end{equation}
the thermodynamic pressure for the system $\beta P(\rho)=\rho/(1-a\rho)$ \cite{1936_Tonks_The} gives the diffusivity reported in Table \ref{tab:trans_params}. 

\section{Numerical confirmation of the fluctuating hydrodynamics} \label{sec:numeric_verify}
 
We perform three independent tests to numerically validate the fluctuating hydrodynamics description of the models discussed in this \emph{Article}.

The first test compares the numerical integration of the hydrodynamic equation \eqref{eq:fhd_general} without the noise term against the evolution of the average density obtained from simulations of the microscopic stochastic dynamics. For the microscopic dynamics, we use the Gillespie algorithm for the lattice models, while for the Langevin particles we integrate the stochastic equation on a periodic geometry. The hydrodynamic equation is solved numerically using the standard Fourier pseudo-spectral method \cite{2000_Trefethen_Spectral}, with nonlinear terms evaluated in real space using the $2/3$ dealiasing rule \cite{2000_Trefethen_Spectral}. Time integration is performed in Fourier space using a standard third-order Runge–Kutta (RK3) method with time step $\mathrm{d}t = 10^{-7}$.

Our second numerical test confirms statistics of macroscopic observables predicted using the fluctuating hydrodynamics equation \eqref{eq:fhd_general}. We consider two well-known observables: the time-integrated current $Q_T$ and the position of a tagged particle $X_T$ measured over a time duration $T$. On an infinite one-dimensional line with initial uniform density profile, $Q_T$ is the net particle flux measured over a duration $T$ across the origin, while $X_T$ is the net displacement of a particle in time $T$. For a single-file diffusive system, where particles preserve their rank due to hard-core constraints, the first two non-trivial cumulants of $Q_T$ and $X_T$ were derived using the fluctuating hydrodynamics description \eqref{eq:fhd_general}. For a system with uniform average density $\bar{\rho}$, the mean $\langle Q_T \rangle$ and $\langle X_T \rangle$ are zero, while their variance is \cite{2012_Krapivsky_Fluctuations,2014_Krapivsky_Large,2015_Krapivsky_Tagged}
\begin{equation}\label{eq:var_current}
\left<Q_T^2\right>_c\simeq\frac{\sigma(\bar{\rho})}{\sqrt{\pi D(\bar{\rho})}}\sqrt{T}\qquad \textrm{and}\qquad 
\left<X_T^2\right>_c\simeq\frac{\sigma(\bar{\rho})}{\bar{\rho}^2\sqrt{\pi D(\bar{\rho})}}\sqrt{T},
\end{equation}
for large $T$. A list of explicit results for the models considered in this \emph{Article} are given in Table \ref{tab:variance_formulae}.

\begin{table}
\centering
\begin{tabular}{|c|c|c|}
\hline
The microscopic model & $\left<Q_T^2\right>_c\big/\sqrt{T}$ & $\left<X_T^2\right>_c\big/\sqrt{T}$ \\
\hline
Symmetric Simple Double Exclusion Process & $\displaystyle{\frac{2\bar{\rho}(1-2\bar{\rho})}{\sqrt{\pi}}}$ & $\displaystyle{\frac{2(1-2\bar{\rho})}{\sqrt{\pi}\bar{\rho}}}$\\[10pt]
Symmetric Simple Triple Exclusion Process & $\displaystyle{\frac{2\bar{\rho}(1-3\bar{\rho})}{\sqrt{\pi}}}$ & $\displaystyle{\frac{2(1-3\bar{\rho})}{\sqrt{\pi}\bar{\rho}}}$\\[10pt]
Symmetric Simple Multiple Exclusion Process & $\displaystyle{\frac{2\bar{\rho}(1-M\bar{\rho})}{\sqrt{\pi}}}$ & $\displaystyle{\frac{2(1-M\bar{\rho})}{\sqrt{\pi}\bar{\rho}}}$\\[10pt]
Brownian hard rods & $\displaystyle{\frac{2\bar{\rho}(1-a\bar{\rho})}{\sqrt{\pi}}}$ & $\displaystyle{\frac{2(1-a\bar{\rho})}{\sqrt{\pi}\bar{\rho}}}$\\[10pt]
Short-range interacting Brownian particles & $\displaystyle{\frac{2\bar{\rho}}{\sqrt{\pi\beta P'(\bar{\rho})}}}$ & $\displaystyle{\frac{2}{\sqrt{\pi\beta P'(\bar{\rho})}\bar{\rho}}}$\\
\hline
\end{tabular}
\caption{Variance of time-integrated current and tagged-particle displacement on an infinite line with uniform average density $\bar{\rho}$.}
\label{tab:variance_formulae}
\end{table}

Explicit expressions for higher-order cumulants of $Q_T$ and $X_T$ beyond fourth order are generally not available \cite{2024_Grabsch_Tracer}. An important exception is the SSEP, for which the full large-deviation statistics have been computed, allowing a complete validation of the predictions of fluctuating hydrodynamics \cite{2007_Derrida_Non,2009_Derrida_Current,2017_Imamura_Large}. For other models, such as the SSMEP or generic Langevin gases, analogous large-deviation results are not known.

For the Symmetric Simple Partial Exclusion Process (SSPEP), the transport coefficients are related to those of the SSEP by a simple rescaling, which enables an explicit determination of the large-deviation statistics in this case as well. Our third numerical test therefore focuses on validating the large-deviation statistics of $Q_T$ in the SSPEP using a rare-event simulation method. Details of these numerical tests are presented below.

\subsection{Symmetric Simple Multiple Exclusion Process}
\begin{figure}[t]
\centering
\begin{subfigure}{0.52\textwidth}
\centering
\includegraphics[width=\linewidth]{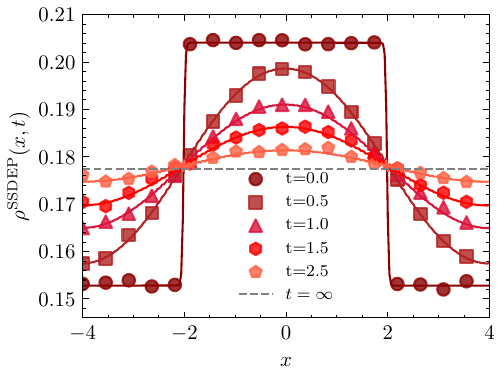}
\caption{\textbf{Hydrodynamic evolution}\label{fig:hydro_evolve_double}}
\end{subfigure}
\\
\begin{subfigure}{0.48\textwidth}
\centering
\includegraphics[width=\linewidth]{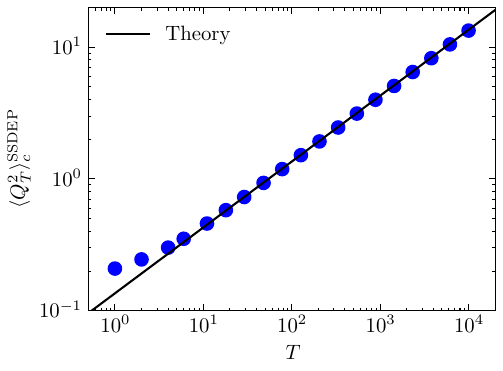}
\caption{\textbf{Variance of the current}\label{fig:current_double}}
\end{subfigure}
\begin{subfigure}{0.48\textwidth}
\centering
\includegraphics[width=\linewidth]{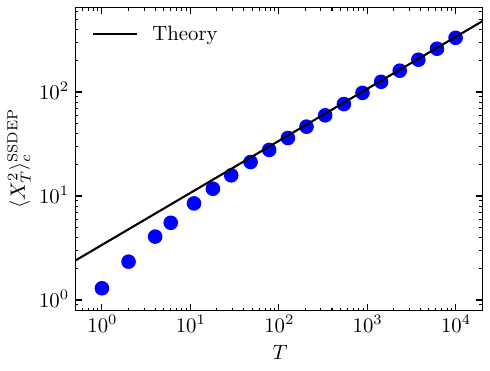}
\caption{\textbf{Variance of the tracer-position}\label{fig:tracer_double}}
\end{subfigure}
\caption{\textbf{Symmetric Simple Double Exclusion Process}: (a) Time evolution of the hydrodynamic density $\rho(x,t)$ starting from a step profile $\rho(x,0)$. The solid lines represents the noiseless hydrodynamic evolution of \eqref{eq:fhd_general} and markers represent the numerical simulations of the microscopic dynamics on a ring of size $L=512$. (b) and (c), the solid line represent the theoretical results for the variance of current and tracer position in an infinite geometry as given in Table \ref{tab:variance_formulae} respectively, while the points represent the numerical simulation. Parameters: $\rho=0.1$, and $10^6$ realizations.
\label{fig:ssdep} 
}
\end{figure}

\begin{figure}[t]
\centering
\begin{subfigure}{0.52\textwidth}
\centering
\includegraphics[width=\linewidth]{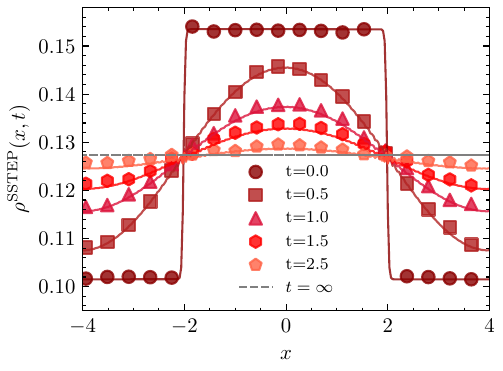}
\caption{\textbf{Hydrodynamic evolution}\label{fig:hydro_evolve_triple}}
\end{subfigure}
\\
\begin{subfigure}{0.48\textwidth}
\centering
\includegraphics[width=\linewidth]{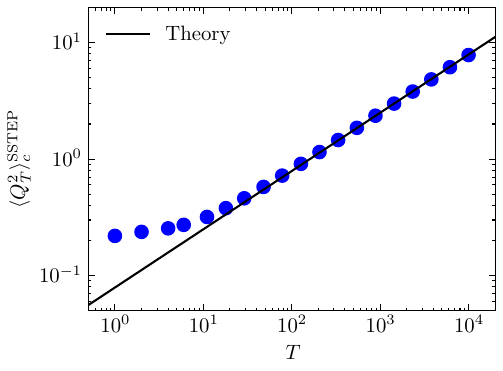}
\caption{\textbf{Variance of the current}\label{fig:current_triple}}
\end{subfigure}
\begin{subfigure}{0.48\textwidth}
\centering
\includegraphics[width=\linewidth]{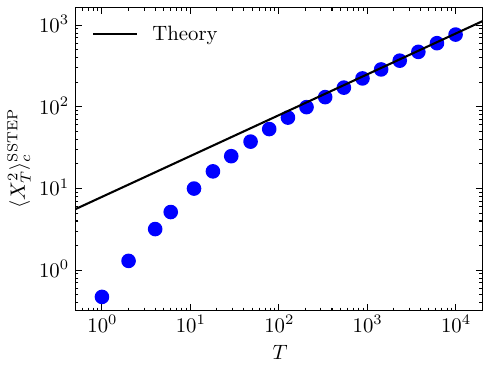}
\caption{\textbf{Variance of the tracer-position}\label{fig:tracer_triple}}
\end{subfigure}
\caption{\textbf{Symmetric Simple Triple Exclusion Process}: (a) Time evolution of the hydrodynamic density $\rho(x,t)$ starting from a step profile $\rho(x,0)$. The solid lines represents the noiseless hydrodynamic evolution of \eqref{eq:fhd_general} and markers represent the numerical simulations of the microscopic dynamics on a ring of size $L=512$. (b) and (c), the solid line represent the theoretical results for the variance of current and tracer position in an infinite geometry as given in Table \ref{tab:variance_formulae} respectively, while the points represent the numerical simulation. Parameters: $\rho=0.1$, and $10^6$ realizations.
\label{fig:sstep} 
}
\end{figure}

We examine the two models of SSMEP—the SSDEP $(M=2)$ and the SSTEP $(M=3)$—to validate their hydrodynamics. Specifically, we compare the evolution of the noiseless hydrodynamic equation \eqref{eq:fhd_general} with microscopic Monte-Carlo simulations on a periodic lattice of size $L = 512$. The initial condition is a step density profile with densities $\rho_a = 0.2$ and $\rho_b = 0.1$ on the two halves of the system. In the microscopic simulations, we measure the time evolution of the averaged occupation numbers $\langle n_i(t)\rangle$ and compare them with the coarse-grained density field $\rho(x,t)$ obtained via the mesoscopic averaging defined in Eq.~\eqref{rhoi_rho}. The microscopic simulations agree well with the noiseless hydrodynamics when using $\ell = 64$ ($\ell $ is the mesoscopic length in \eqref{rhoi_rho} with $0 \ll \ell \ll L$.), as shown in Fig.~\ref{fig:hydro_evolve_double} and Fig.~\ref{fig:hydro_evolve_triple} for the SSDEP and SSTEP respectively.

Next, to study the variance of the time-integrated current and the tracer position, we consider the infinite-geometry case. The initial configuration is sampled from the steady state: we prepare a Bernoulli product measure at density $\rho$, with a length constraint and a designated tracer particle fixed at the origin $x = 0$. The numerical results for the variance of the current $Q_t$ are presented in Fig.~\ref{fig:current_double} and Fig.~\ref{fig:current_triple} for the SSDEP and SSTEP respectively, and show excellent agreement with the theoretical predictions summarized in Table~\ref{tab:variance_formulae}. We also numerically verify the tracer-position variance in Fig.~\ref{fig:tracer_double} and Fig.~\ref{fig:tracer_triple} for the two models.

\subsection{Brownian gas with WCA interactions}
\begin{figure}[th!]
\centering
\begin{subfigure}{0.52\textwidth}
\centering
\includegraphics[width=\linewidth]{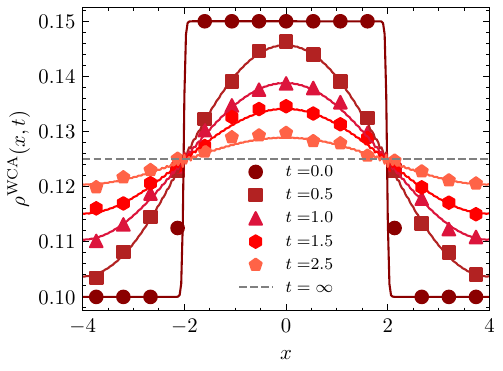}
\caption{\textbf{Hydrodynamic evolution}\label{fig:wca}}
\end{subfigure}
\\
\begin{subfigure}{0.48\textwidth}
\centering
\includegraphics[width=\linewidth]{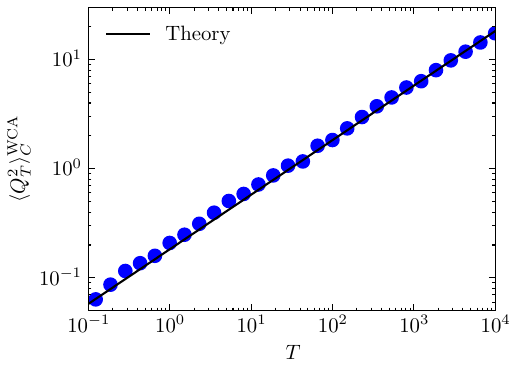}
\caption{\textbf{Variance of the current}\label{fig:Current_variance_BMs_WCA}}
\end{subfigure}
\begin{subfigure}{0.48\textwidth}
\centering
\includegraphics[width=\linewidth]{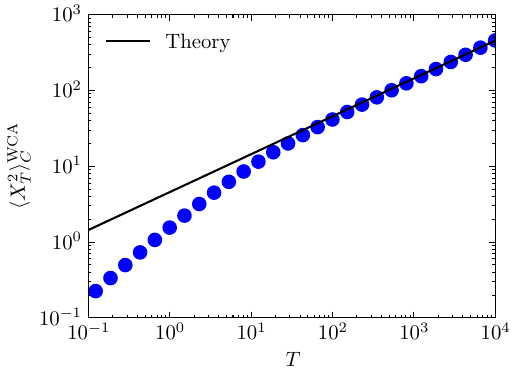}
\caption{\textbf{Variance of the tracer-position}\label{fig:Tracer_variance__BMs_WCA}}
\end{subfigure}
\caption{\textbf{Brownian gas with WCA interactions}: (a) Time evolution of the hydrodynamic density $\rho(x,t)$ starting from a step profile $\rho(x,0)$. The solid lines represent the noiseless hydrodynamics evolution of \eqref{eq:fhd_general} and markers represent the numerical simulations of the microscopic dynamics given by \eqref{eq:Langevian_eq} with \eqref{eq:WCA_pot} on a ring of size $L=1200$ and averaging over $4000$ realizations. (b) and (c), the solid line represent the theoretical result for the variance of current and tracer position as given in Table \ref{tab:variance_formulae} respectively, while the points represent the numerical simulation for $\rho=0.2$, $L=5000$, and $1212$ realizations. Remaining simulation parameters (a)-(c): $\mu=1$, $D_0 = 1$, $\beta=1$, $\epsilon=1$, $a=1$.\label{fig:placeholder}}
\end{figure}

To validate the hydrodynamic equation \eqref{eq:brownian_fhd} for the continuum models, we study a Brownian gas with Weeks-Chandler-Andersen(WCA) interaction \cite{1971_Weeks_Role},
\begin{equation}\label{eq:WCA_pot}
V(r)=
\begin{cases}
4\varepsilon\left[\left(\dfrac{a}{r}\right)^{12}-\left(\dfrac{a}{r}\right)^6 \right]+\varepsilon & \text{for }\left|r\right|<2^{1/6}a\\[6pt]
0 & \text{for }\left|r\right|\geq2^{1/6}a
\end{cases}
\end{equation}
where $\varepsilon$ is the strength of the repulsion and $a$ denotes the characteristic interaction length. The exact equation of state for the pressure is unknown, but it can be approximated at low-to-moderate densities using the virial expansion. For For $\beta = 1, \epsilon =1$ and $a=1$, it reads as \cite{2013_Hansen_Liquids,2025_Grabsch_Exact}


\begin{equation}
P(\rho) = \rho + 1.01561 \rho^2 +1.02977 \rho^3 \,,
\end{equation}
so the diffusivity $D(\rho)= \beta P'(\rho)$ becomes
\begin{equation}
D(\rho) = 1 + 2.03122 \rho + 3.08931 \rho^2 \,.
\end{equation}
Using these transport coefficients, we first numerically verify the noiseless hydrodynamic equation \eqref{eq:fhd_general} against microscopic dynamics given in \eqref{eq:Langevian_eq} with \eqref{eq:WCA_pot}. The microscopic dynamics are integrated with the Euler-Maruyama scheme using a timestep $\mathrm{d}t = 10^{-4}$ on a ring of length $L=1200$. Initial configurations are deterministic step profiles with $\rho_a = 0.15$ and $\rho_b = 0.1$. The coarse-grained microscopic density matches the hydrodynamic prediction (with mesoscopic length $\ell = 150$) as show in Fig. \ref{fig:wca}. 

We also verify the theoretical expressions for the variance of the time-integrated current and of a tracer particle. For these confirmations we simulated a ring of length $L= 5000$ at global density $\rho = 0.2$ using a timestep $\rm d $$t = 10^{-4}$ . Initial configurations were sampled from the steady state. The long-time variances of the current and tracer measured in microscopic simulations closely follow the theoretical predictions, as summarized in Table \ref{tab:variance_formulae} and shown in Fig. \ref{fig:Current_variance_BMs_WCA} and Fig. \ref{fig:Tracer_variance__BMs_WCA} respectively. 

\subsection{Symmetric Simple Partial Exclusion Process}
Using the relation between the transport coefficients of the SSPEP and the SSEP, we show that for the former, the scaled cumulant-generating function of the time-integrated current
\begin{equation}
\mu(\lambda,\rho_a,\rho_b)=\lim_{T\to\infty}\frac{\ln{\left<\mathrm{e}^{\lambda Q_T}\right>}}{\sqrt{T}} \,,
\end{equation}
is given via a simple relation
\begin{subequations}
\begin{equation}\label{eq:scgf}
\mu_{\text{SSPEP}}(\lambda,\rho_a,\rho_b)=N^{3/2}\mu_{\text{SSEP}}\!\left(\lambda,\frac{\rho_a}{N},\frac{\rho_b}{N}\right) \,,
\end{equation}
with the scaled cumulant-generating function in the SSEP \cite{2009_Derrida_Current,2022_Mallick_Exact} given by
\begin{equation}
\mu_{\text{SSEP}}\!\left(\lambda,\rho_a,\rho_b\right)=\frac{1}{\pi}\int_{-\infty}^{\infty}\mathrm{d}k\,\ln{\left\{1+\left[(\mathrm{e}^{\lambda}-1)\rho_a(1-\rho_b)+(\mathrm{e}^{-\lambda}-1)\rho_b(1-\rho_a)\right]\mathrm{e}^{-k^2}\right\}} \,,
\end{equation}
\end{subequations}
and where $N$ is the maximal occupancy of each site in the SSPEP (see Fig. \ref{fig:sshep_scheme} for the $N=2$ case). Similarly, the large deviations for the density profile in the non-equilibrium steady state of a boundary-driven SSPEP can be related to that of the SSEP \cite{2008_Tailleur_Mapping,2025_Saha_Large}. These results are remarkable given that the microscopic dynamics of the SSPEP is not integrable via the Bethe ansatz. Nevertheless, its fluctuating hydrodynamic description remains tractable, underscoring the versatility and robustness of the hydrodynamic framework.

We confirm the result given in \eqref{eq:scgf} by rare event numerical simulations based on the continuous-time cloning algorithm \cite{2006_Giardinà_Direct, 2007_Lecomte_Numerical, 2019_Perez_Sampling}. To obtain the scaled cumulant-generating function of the current for the SSPEP for $N=2$ on an infinite lattice, we performed numerical simulations on a finite lattice of length \(2L+1\) with reflecting boundary conditions at \(-L\) and \(+L\). Initially, the lattice is populated with densities \(\rho_a\) and \(\rho_b\) to the left and right of the origin, respectively. The net current is measured across the origin over a time period \(T \ll L^2\), ensuring that the effects of the reflecting boundaries remained negligible.

\begin{figure}[t]
\centering
\includegraphics[width=0.6\linewidth]{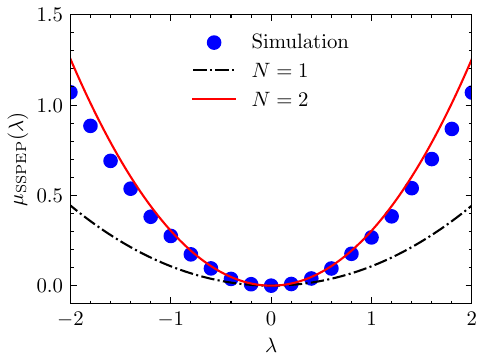}
\caption{\textbf{Scaled cumulant-generating function of the current in the SSPEP with $N=2$}: The solid red line represents the theoretical result of \eqref{eq:scgf} for $\rho_{a(b)}=0.5$, with blue circles representing the corresponding simulation result obtained by the cloning algorithm. The results closely match in the range $-1\leq\lambda\leq1$. The noticeable deviations from the theoretical results near large $\lambda$ is due to finite size effects. For comparison, the scaled cumulant-generating function of the current in the SSEP ($N=1$)\cite{2009_Derrida_Current,2022_Mallick_Exact} is shown in black dashed line.}


\label{fig:SSPEP_inf_cgf}
\end{figure}

The scaled cumulant-generating function of the current is obtained on a finite lattice with $L=75$ up to a measurement time of $T=500$, using a clone population of \(N_c=40000\) and averaging over $10$ independent samples. Figure \ref{fig:SSPEP_inf_cgf} successfully compares these results with the theoretical expectation for $\rho_a=\rho_b=1/2$, in a range of parameters $\lambda$.

\section{Conclusion}\label{sec:concl}

In this \emph{Article}, we present a systematic coarse-graining approach for the fluctuating hydrodynamics description in a class of stochastic lattice-gas models with exclusion interactions. Our approach is based on a path-integral formulation of the microscopic stochastic dynamics and subsequent coarse-graining for obtaining the Martin-Siggia-Rose-Janssen-De Dominicis (MSRJD) path-integral of the coarse-grained density field. The corresponding stochastic differential equation is the fluctuating hydrodynamics describing the large-scale evolution of the lattice-gas. In this coarse-grained description all the microscopic details of the underlying dynamics enter only through the two transport coefficients: the diffusivity and mobility.

In the later half of the \emph{Article} we extend our coarse-graining approach to the continuum setting of the Dean–Kawasaki equation describing Brownian particles interacting through a short-range two-body interaction potential. The coarse-graining recovers the fluctuating hydrodynamics which was mostly argued earlier on phenomenological ground and sets the basis of the Macroscopic Fluctuation Theory. Our analysis shows how the interaction potential in the Dean-Kawasaki equation renormalizes the bare diffusivity, expressing it in terms of the thermodynamic pressure, while the mobility remains a linear function of the coarse-grained density.

Note that the coarse-graining of the Dean-Kawasaki equation could be equivalently performed at the level of the stochastic partial differential equation. However, the path integral approach provides a unifying methodology for both lattice and off-lattice dynamics.

In this \emph{Article} we have deliberately not addressed the role of external potentials. 
For a smoothly varying potential of the form \(U(x)=u(x/\ell)\), its effect can be incorporated 
within linear response theory, leading to the fluctuating hydrodynamics
\begin{equation}\label{eq:fhd_general_drift}
\partial_t \rho
= \partial_x\!\left( D(\rho)\,\partial_x \rho 
+ \sigma(\rho)\,u'(x) \right)
+ \frac{1}{\sqrt{\ell}}\,\partial_x\!\left( \sqrt{\sigma(\rho)}\,\eta \right).
\end{equation}
In contrast, a rapidly varying external potential is expected to renormalize the bare transport 
coefficients. See analogous results in \cite{1962_Lifson_On,1983_Derrida_Velocity}. Explicitly constructing the corresponding fluctuating hydrodynamics within our framework remains an interesting direction for future work.

Building a quantitative fluctuating hydrodynamics for larger class of dynamics, such as ballistic transport \cite{1991_Krug_Boundary,2014_Spohn_Nonlinear} or integrable dynamics \cite{2016_Bertini_Transport,2016_Castro_Emergent} remains a challenge. The coarse-graining scheme presented in this \emph{Article} may prove useful in this endeavor. For active dynamics, which are ballistic at short scales and diffusive beyond, our method proves useful in constructing a bottom-up fluctuating hydrodynamics for standard active matter models \cite{2013_Marchetti_Hydrodynamics,2024_Dinelli_Fluctuating} such as the active Brownian particles (ABP) \cite{2012_Fily_Athermal} and active Ornstein-Uhlenbeck particles (AOUP) \cite{2014_Szamel_Self}. A detailed discussion will be presented elsewhere.

Our coarse-graining method crucially depends on the underlying gradient structure \cite{2025_Derrida_Lectures} of the particle current at the microscopic scales. For gradient dynamics in higher dimensions, our method straightforwardly extends. A natural question is to explore whether our approach could be extended for non-gradient models \cite{1994_Kipnis_Hydrodynamical,2001_Landim_Hydrodynamic,2014_Arita_Generalized}, particularly recovering the variational formula \`a la Spohn and Varadhan \cite{1991_Spohn_Large,1983_Spohn_Long,1988_Guo_Nonlinear,2017_Arita_Variational} for the diffusivity. We emphasize that in our coarse-graining approach, mobility can still be computed without the gradient structure (see Appendix \ref{sec:explicit}).

\section*{Acknowledgments}
We thank Bernard Derrida for several invaluable discussions during the course of this project, particularly for seeding the idea of $M$-site exclusion process for studying the hard-rods problem.

\paragraph{Funding information}
SS, SJ and TS acknowledge financial support of the Department of Atomic Energy, Government of India, under Project Identification No. RTI 4002. SS gratefully acknowledges the Infosys–TIFR Leading Edge Travel Grant for financial support during participation in the conference StatPhys29 and the Satellite Meeting NESP-2025, where part of this work was carried out. TS thanks the support from the International Research Project (IRP) titled `Classical and quantum dynamics in out of equilibrium systems' by CNRS, France. TADP thanks TIFR, Mumbai for hosting visits which helped initiate the project.

\begin{appendix}
\numberwithin{equation}{section}

\section{Further details about coarse-graining}\label{sec:explicit}
We present here a detailed discussion of our approach for fluctuating hydrodynamics for diffusive lattice gases aiming to clarify the various steps taken in Sec.~\ref{sec:lattice} of the main text. We particularly emphasize that the response field can be treated as a smooth field and that the local equilibrium measure enters as a consequence of the law of large numbers.

We follow the notations used in Section \ref{sec:lattice}. We denote the occupation number at site $i$ as $n_i$. Due to particle conservation, we write the microscopic continuity equation $n_i(\tau+\mathrm{d}\tau)-n_i(\tau)=Y_i(\tau)-Y_{i+1}(\tau)$ with
\begin{equation}\label{eq:rates_app}
Y_i(\tau)=
\begin{cases}
+1, & \text{with prob. } r_i^+(\tau)\mathrm{d}\tau\,,\\
-1, & \text{with prob. } r_i^-(\tau)\mathrm{d}\tau\,,\\
0, & \text{with prob. } 1-r_i^+(\tau)\mathrm{d}\tau-r_i^-(\tau)\mathrm{d}\tau \,,
\end{cases}
\end{equation}
where $r_i^{\pm}(\tau)$ are local functions of the occupation numbers at time $\tau$, as in the SSDEP \eqref{eq:right_left_rate_defn}, SSTEP \eqref{eq:sstep_rgt_lft_rate} and SSPEP \eqref{eq:sspep_rgt_lft_rate} discussed earlier. We introduce a macroscopic length-scale $\ell$ and a mesoscopic coarse-graining length $\Lambda$ (with eventually $\ell\to\infty$, $\Lambda\to\infty$ and $\ell/\Lambda\to\infty$). We define the coarse-grained density field as
\begin{equation*}
\rho(x,\tau) = \sum_j K_\Lambda(j)\,n_{i+j}(\tau)\quad;\quad i=\ell x\,,
\end{equation*}
where the kernel $K_\Lambda$ is nonzero on scales $O(\Lambda)$ and is such that
\begin{equation*}
\sum_j K_\Lambda(j) = 1\,.
\end{equation*}
In the following, we take $x\in[0,1]$, $\mathrm{d}x=1/\ell $ and we use the notation
\begin{equation*}
\sum_x \mathrm{d}x f(x) = \sum_{i=1}^\ell \frac{1}{\ell }f\left(\frac{i}{\ell }\right)\,.
\end{equation*}
The dynamics of the set of occupation numbers $\{n_i\}$ is Markovian. Here, we show that, on large scales, the dynamics of the coarse-grained field $\rho(x,\tau)$ also becomes Markovian. To proceed, we introduce
\begin{equation*}
\Delta_x = \frac{\rho(x,\tau+d\tau)-\rho(x,\tau)}{\mathrm{d} \tau} \,,
\end{equation*}
and we evaluate the conditional probability $\mathcal{P}(\{\Delta_x\}|\{n_i(\tau)\})$ of the coarse-grained density field at time $\tau+\mathrm{d}\tau$ knowing the occupation numbers at time $\tau$. We introduce the notation $\left\langle \dots \right\rangle_0$ for the averages over the initial state at time $\tau$, and denote $\bar{\rho}(x) \equiv \left\langle n_{\ell x}(t) \right\rangle_0$. We assume that the function $\bar{\rho}(x)$ varies smoothly over scales of order $O(1)$, meaning that the average density varies over scales of order $O(\ell )$. We further assume that the correlation length $\xi$ in the initial state is small compared to the mesoscopic scale, that is $\xi \ll \Lambda$. It then follows from the law of large numbers that, in the initial state,
\begin{equation}\label{eq:LLN_density}
\rho(x,\tau) \simeq \bar{\rho}(x) \,,
\end{equation}
up to small corrections scaling as $\Lambda^{-1/2}$. We further assume that the initial state is in a local equilibrium, meaning that, for any local function of the occupation numbers, the initial state averages follow
\begin{equation*}
\left\langle f(n_{\ell x}) \right\rangle_0 = \left\langle f(n) \right\rangle^{\text{eq}}_{\bar{\rho}(x)} 
\end{equation*}
where $\left\langle \dots \right\rangle^{\rm eq}_{\bar{\rho}(x)}$ denotes a steady-state average in a homogeneous system with mean density $\bar{\rho}(x)$. 

To construct the probability $\mathcal{P}(\{\Delta_x\}|\{n_i(\tau)\})$, we follow \cite{2006_Andreanov_Exact,2007_Lefevre_Dynamics,2024_Saha_Large,2025_Mukherjee_Hydrodynamics} and the steps outlined in Sec.~\ref{sec:lattice}. We obtain, 
\begin{align*}
\mathcal{P}(\{\Delta_x\}|\{n_i(\tau)\}) & = \int \prod_x\frac{\ell \mathrm{d} x\mathrm{d} \tau}{2\pi}\mathrm{d} \widehat{\rho}_x\exp\left(i\,\mathrm{d} \tau\,\ell \sum_x \mathrm{d} x \widehat{\rho}_x \Delta_x\right) \\
& \quad\;\left\langle \exp\left(-i\ell \sum_x \mathrm{d} x \widehat{\rho}_x \sum_j K_\Lambda(j)\left[Y_{\ell x+j}(\tau)-Y_{\ell x + j + 1}(\tau)\right]\right)\right\rangle \,.
\end{align*}
We now perform a summation by parts and, neglecting boundary terms, get
\begin{equation*}
-i\ell \sum_x \mathrm{d}x \widehat{\rho}_x \sum_j K_\Lambda(j)\left[Y_{\ell x+j}(\tau)-Y_{\ell x + j + 1}(\tau)\right] = -i\sum_x \mathrm{d}x \partial_x\widehat{\rho} \sum_j K_\Lambda(j)Y_{\ell x + j}(\tau) \,,
\end{equation*}
with the notation
\begin{equation*}
\partial_x\widehat{\rho} = \ell \left(\widehat{\rho}_{x+1/\ell }-\widehat{\rho}_x\right) \,.
\end{equation*}
Using (\ref{eq:rates_app}), we can average over the variables $Y_{i}(\tau)$ and obtain, to first order in $\mathrm{d}\tau$
\begin{align*}
& \left\langle \exp\left(-i\ell \sum_x \mathrm{d}x \widehat{\rho}_x \sum_j K_\Lambda(j)\left[Y_{\ell x+j}(\tau)-Y_{\ell x + j + 1}(\tau)\right]\right)\right\rangle \\
& = \exp\Bigg(\mathrm{d}\tau\sum_k r_k^+(\tau)\left[\exp\left(-i\sum_x \mathrm{d}x \, \partial_x\widehat{\rho} \, K_\Lambda(k - \ell x)\right)-1\right] \\
& \quad\quad\quad\;+ \mathrm{d}\tau\sum_k r_k^-(\tau) \left[\exp\left(i\sum_x \mathrm{d}x \, \partial_x\widehat{\rho} \, K_\Lambda(k - \ell x)\right)-1\right]\Bigg) \,,
\end{align*}
We now note that the sums inside the internal exponentials scale as $O(\ell ^{-1})$. In fact, the kernel $K_\Lambda(k-\ell x)$ constrains the sums over $x$ to run over $O(\Lambda)$ terms, while the kernel itself is of order $O(\Lambda^{-1})$. Therefore, $\mathrm{d}x = O(\ell ^{-1})$ sets the scale of these sums and we can expand the internal exponentials up to the order $O(\ell ^{-2})$. With the sum over $k$, this means expanding the action up to order $O(\mathrm{d}\tau \ell ^{-1})$ which becomes $O(\ell )$ after a diffusive rescaling of time $\tau = \ell ^2t$. To that order, we get
\begin{align}
& \left\langle \exp\left(-i\ell \sum_x \mathrm{d}x \widehat{\rho}_x \sum_j K_\Lambda(j)\left[Y_{\ell x+j}(\tau)-Y_{\ell x + j + 1}(\tau)\right]\right)\right\rangle \nonumber\\
=&\exp\left(-i \mathrm{d}t \sum_x\, \mathrm{d}x\, \partial_x\widehat{\rho} \sum_k K_\Lambda(k) \left(r_{\ell x+k}^+(\tau) - r_{\ell x+k}^-(\tau)\right)\right) \nonumber\\
& \exp\left( - \mathrm{d}\tau \sum_x  \mathrm{d}x \, \partial_x \widehat{\rho}  \sum_k K_\Lambda(k) \left[\sum_{x'} \, \mathrm{d}x'  K_\Lambda(k + \ell (x-x')) \, \partial_{x'}\widehat{\rho}\right] \frac{r_{\ell x+k}^+(\tau) + r_{\ell x+k}^-(\tau)}{2} \right).\label{eq:action_LLN}
\end{align}
We start by discussing the second term, which is quadratic in the response field. By virtue of the power counting argument given above, we only need to consider it to leading order in $\ell $, which leads to
\begin{equation*}
\sum_{x'} \, \mathrm{d}x'  K_\Lambda(k + \ell (x-x')) \, \partial_{x'}\widehat{\rho} = \frac{1}{\ell } \sum_{k'} K_\Lambda(k') \partial_{x}\widehat{\rho}\left(x+k/\ell -k'/\ell \right) \simeq \frac{1}{\ell } \partial_{x}\widehat{\rho} \,,
\end{equation*}
because both $k$ and $k'$ are of order $\Lambda$. Thus we are left with
\begin{align*}
& - \mathrm{d}t \sum_x  \mathrm{d}x \, \partial_x \widehat{\rho}  \sum_k K_\Lambda(k) \left[\sum_{x'} \, \mathrm{d}x'  K_\Lambda(k + \ell (x-x')) \, \partial_{x'}\widehat{\rho}\right] \frac{r_{\ell x+k}^+(\tau) + r_{\ell x+k}^-(\tau)}{2}\nonumber\\
\simeq&-\frac{\mathrm{d}\tau}{\ell} \sum_x  \mathrm{d}x \left(\partial_x \widehat{\rho}\right)^2 \sum_k K_\Lambda(k) \frac{r_{\ell x+k}^+(\tau) + r_{\ell x+k}^-(\tau)}{2} \,.
\end{align*}
As in (\ref{eq:LLN_density}), we can use the law of large numbers and the local equilibrium hypothesis to obtain to leading order in the hydrodynamic limit,
\begin{equation*}
\sum_k K_\Lambda(k) \frac{r_{\ell x+k}^+(\tau) + r_{\ell x+k}^-(\tau)}{2} \simeq \left\langle  \frac{r^+ + r^-}{2} \right\rangle^{\rm eq}_{\rho(x,\tau)} \,.
\end{equation*}
This sets the noise amplitude in the fluctuating hydrodynamics. Deriving the deterministic part is more subtle and relies on a next-to-leading order analysis of the terms linear in the response field appearing in (\ref{eq:action_LLN}), as these are formally of order $O(\ell ^{-1})$. A major simplification arises, however, for gradient systems, where there exists a function $h_i$ of the occupation numbers such that the average current can be written as a discrete gradient, 
\begin{equation*}
r_{\ell x+k}^+(\tau) - r_{\ell x+k}^-(\tau) = h_{\ell x + k + 1}(\tau) - h_{\ell x + k}(\tau) \,,
\end{equation*}
as studied in this \emph{Article}. Then, an additional summation by parts can be performed
\begin{align*}
& -i \mathrm{d}t \sum_x\, \mathrm{d}x\, \partial_x\widehat{\rho} \sum_k K_\Lambda(k) \left(r_{\ell x+k}^+(\tau) - r_{\ell x+k}^-(\tau)\right) \\
=&i \mathrm{d}t \sum_x\, \mathrm{d}x\, \sum_k K_\Lambda(k) \left(\partial_{x}\widehat{\rho}(x+1/\ell ) - \partial_x\widehat{\rho}\right)  h_{\ell x + k}(\tau) \\
=&\frac{i \mathrm{d}t}{\ell} \sum_x\, \mathrm{d}x\,\partial^2_{x}\widehat{\rho} \, \left\langle h \right\rangle^{\rm eq}_{\rho(x,\tau)}
\end{align*}
where the last line is obtained using the law of large numbers and the local equilibrium hypothesis. To the considered order, the resulting transition probability between $\rho(x,\tau+\mathrm{d}\tau)$ and $\rho(x,\tau)$ thus only depends on $\rho(x,\tau)$ and not upon finer details of the initial state. If we assume that the local equilibrium hypothesis is preserved under time evolution, this means that the dynamics of the coarse-grained density field is effectively Markovian. Therefore, after a diffusive rescaling of time, we get the path integral formulation 
\begin{equation*}
\mathcal{P}(\rho(x,T)|\rho(x,0) = \int_{\rho(x,0)}^{\rho(x,T)}[\mathcal{D}\rho][\mathcal{D}\widehat{\rho}][\mathcal{D}\rho] \,\mathrm{e}^{-\ell \mathcal{S}[\rho,\widehat{\rho}]}
\end{equation*}
with the action
\begin{equation*}
\mathcal{S}[\rho,\widehat{\rho}] = \int_0^T \mathrm{d}t \int \mathrm{d}x \left[\widehat{\rho}\partial_t\rho + \partial_x\widehat{\rho} \,\partial_x\! \left\langle h \right\rangle^{\text{eq}}_{\rho} - \left(\partial_x \widehat{\rho}\right)^2 \left\langle  \frac{r^+ + r^-}{2} \right\rangle^{\text{eq}}_{\rho}\right] \,.
\end{equation*}
Comparing with \eqref{eq:trans_prob_general}, we see that
\begin{equation*}
\partial_x\big<h\big>_{\rho}^{\text{eq}}=D(\rho)\partial_x\rho\;\;\text{and}\;\;\left\langle\frac{r^+ + r^-}{2}\right\rangle^{\text{eq}}_{\rho}=\frac{\sigma(\rho)}{2}.
\end{equation*}

\end{appendix}

\bibliography{references.bib}

\end{document}